\documentclass[12pt,preprint]{aastex}

\begin{document}

 \def\Mdot{\hbox{$\dot {M}$}}
 \def\Zdot{\hbox{$\dot {Z}$}}
 \def\Rsun{\hbox{\it R$_\odot$}}
 \def\Rstar{\hbox{\it R$_*$}}
 \def\Lsun{\hbox{\it L$_\odot$}}
 \def\Lstar{\hbox{\it L$_*$}}
 \def\Msun{\hbox{\it M$_\odot$}}
 \def\Msunyr{\hbox{\it M$_\odot\,$yr$^{-1}$}}
 \def\Myr{\hbox{\it Myr}}
 \def\Gyr{\hbox{\it Gyr}}
 \def\Teff{\hbox{\it T$_{\rm eff}$}}
 \def\Vinf{\hbox{$v_\infty$}}
 \def\kms{\hbox{km$\,$s$^{-1}$ }}

\title{High Spectral Resolution Observations of the Massive Stars in the Galactic Center}

\shorttitle{Massive Stars in the GC}

\author{Angelle Tanner\altaffilmark{1},
Donald F. Figer\altaffilmark{2}, Francisco Najarro\altaffilmark{3}, Rolf P. Kudritzki\altaffilmark{4}, Diane Gilmore\altaffilmark{2}, \\
Mark Morris\altaffilmark{5}, E. E. Becklin\altaffilmark{5}, Ian S. McLean\altaffilmark{5}, \\
Andrea M. Gilbert\altaffilmark{6}, James R. Graham\altaffilmark{7}, \\
James E. Larkin\altaffilmark{5}, N. A. Levenson\altaffilmark{8},
Harry I. Teplitz\altaffilmark{9}}

\altaffiltext{1}{Jet Propulsion Lab, 4800 Oak Grove Dr., Pasadena, CA 91106 }
\altaffiltext{2}{Space Telescope Science Institute, 3700 San Martin Drive, Baltimore, MD 21218 }
\altaffiltext{3}{Instituto de Estructura de la Materia, CSIC, Serrano 121, 28006, Madrid, Spain}
\altaffiltext{4}{Institute for Astronomy, University of Hawaii, 2680 Woodlawn Drive, Honolulu, HI 96822}
\altaffiltext{5}{Department of Astronomy and Astrophysics, UCLA, 405 Hilgard Ave., Los Angeles, CA 90095}
\altaffiltext{6}{Max-Planck-Institut für extraterrestrische Physik, Giessenbachstr. 1, 85748 Garching, Germany}
\altaffiltext{7}{Department of Astronomy, University of California, Berkeley, 601 Campbell Hall, Berkeley, CA, 94720-3411}
\altaffiltext{8}{Department of Physics and Astronomy, University of Kentucky, 600 Rose Street, Lexington, Kentucky, 40506-0055}
\altaffiltext{9}{Spitzer Science Center, MS 220-6, California Institute of Technology, Pasadena, CA, 91125}
\authoremail{angelle.tanner@jpl.nasa.gov}

\begin{abstract}

We present high-resolution near-infrared spectra, obtained with the NIRSPEC spectrograph 
on the W. M. Keck II Telescope, of a collection of hot, massive stars 
within the central
25 arcseconds of the Galactic center. We have identified a total of twenty-one emission-line stars, 
seven of which are new radial velocity detections with five of those being classified
as He I emission-line stars for the first time. 
These stars fall into two categories based on their spectral properties: 1) those with
narrow $\lambda\lambda$ 2.112, 2.113 $\micron$ \ion{He}{1} doublet absorption lines, and 
2) those with broad 2.058 $\micron$ \ion{He}{1} emission lines.
These data have the highest spectral resolution ever obtained for these sources 
and, as a result, both components
of the absorption doublet are separately resolved for the first time.
We use these spectral features to measure radial velocities.
The majority of the measured radial velocities have
relative errors of 20 \kms, smaller than those previously obtained with proper-motion or radial
velocity measurements for similar stellar samples in the Galactic center. The radial 
velocities estimated from the
\ion{He}{1} absorption doublet are more robust than those previously estimated from the 
2.058 $\micron$ emission line,
since they do not suffer from confusion due to emission from the surrounding ISM.
Using this velocity information, we agree with previous stellar velocity studies 
that the stars are orbiting in a somewhat coherent 
manner but are not as defined into a disk or disks as previously thought. Finally, 
multi-epoch radial
velocity measurements for IRS 16NE show a change in its velocity presumably due
to an unseen stellar companion.

\end{abstract}


\section{INTRODUCTION}

The dynamic environment created by the presence of a supermassive black hole
at the Galactic center is thought to inhibit the classical process of star formation  
initiated through the spontaneous contraction of
material to densities above the Jeans limit (Morris 1993). As a result, external
influences may be required such as shocks due to cloud collisions,
supernovae, and stellar winds to compress the material prior to star formation.
These methods for star formation may skew the initial mass function (IMF) 
toward the production of massive stars and, indeed,
the stellar population in the central parsec of the galaxy, as well as the nearby Quintuplet and Arches
clusters, contains a significant number of the massive stars.
While the Galactic center would seem to be a hostile environment for recent
star formation, the region is also replete with young stars with the central
parsec hosting a large number of young, massive He I emission-line stars 
(Rieke \& Lebofsky 1987, Lebofsky \& Rieke 1987; Forrest et al. 1987; Krabbe et al. 1991;
Rieke 1993; Blum et al. 1995).

This paradox of recent star formation in an inhospitable environment 
is even more dramatic for the young stars within one arcsecond of the supermassive
black hole (Genzel et al. 1997; Eckart et al. 1999; Figer et al. 2000;
Gezari et al. 2002; Ghez et al. 2003). Many stellar dynamic models have been used to 
try to explain the presence of all stars observed in the central parsec. 
For instance, rather than {\it in situ} star formation, one mechanism proposed to
explain the presence of the young stars involves the in-fall of
a parent supermassive stellar cluster from outside the central parsec due to
dynamical friction. This mechanism would have to explain the observed stellar
populations both within
the central arcsecond and in the surrounding He I emission-line star cluster
(Kim et al. 1999; Portegies-Zwart et al. 2001; Gerhard 2001; Kim et al. 2004).
N-body simulations of the massive cluster's in-fall via dynamical friction,
however, do not result in a large
enough population of He I emission-line stars compared to what is known from observations,
nor do the simulations address
the presence of the stars in the central arcsecond (Gerhard 2001; Kim \& Morris
2003; Paumard et al. 2003; Kim et al. 2004).
Another mechanism which could insert young, 
massive stars into the central parsec is the presence of an intermediate mass black hole
within the in-falling cluster. Such a black hole would act as a
gravitational seed dragging the massive stars along during in-fall 
(Hansen \& Milosavljevi{\'c} 2003; Kim et al. 2004). 
While there is no definitive evidence to date for an intermediate-mass black hole
within the central parsec, one small cluster of stars, IRS 13, is
a prominent X-ray source and is a candidate location for the presence of an IMBH (Maillard et al. 2004). 
Finally, young stars could be inserted into an orbit one arcsecond from Sgr A* through
multi-body interactions from in-falling massive binary systems (Gould \&
Quillen 2003). With the young star problem being key to understanding
star formation within the central parsec, it becomes essential to carry out a complete census of this stellar
population in order to test the star formation models.

Many early-type stars in the central parsec, including the brightest members
of the IRS 16 cluster, have been identified as luminous blue variable (LBV) candidates,
or Wolf-Rayet (WR)
stars, based on the atomic transitions and equivalent widths (EWs) of their observed emission
lines (Krabbe et al. 1995; Blum et al. 1996; Paumard et al. 2001, 2003, hereafter P2001, P2003;
Genzel et al. 2000, hereafter G2000; Tamblyn et al. 1996).
The presence of ionized gas within the central parsec often generates
confusion when classifying stellar spectra requiring
high spatial resolution observations which have only been 
possible within the past few years with 5 to 10
meter class telescopes. The Wolf-Rayet
stars in the central parsec, and in the nearby Quintuplet and Arches clusters, 
including late-type WC and Ofpe/WNL stars, constitute the largest
population of such stars in the Galaxy (Figer et al. 1995; Figer et al. 1996; Cotera et
al. 1996; Figer et al. 1999; Figer et al. 2002; Blum et al. 2001; Genzel et al. 2000).

In this paper, we present new high-spectral resolution near-infrared spectra
of a collection of massive, hot stars within the central parsec of the Galaxy.
Section 2 describes the data set obtained for this work, Section 3 describes
the techniques used to extract information from the spectra, Section 4
reviews the information we can derive from the data, and Section 5 discusses
the implications of these results.

\section{OBSERVATIONS}

\begin{figure}[ht]
\epsscale{1.0}
\plotone{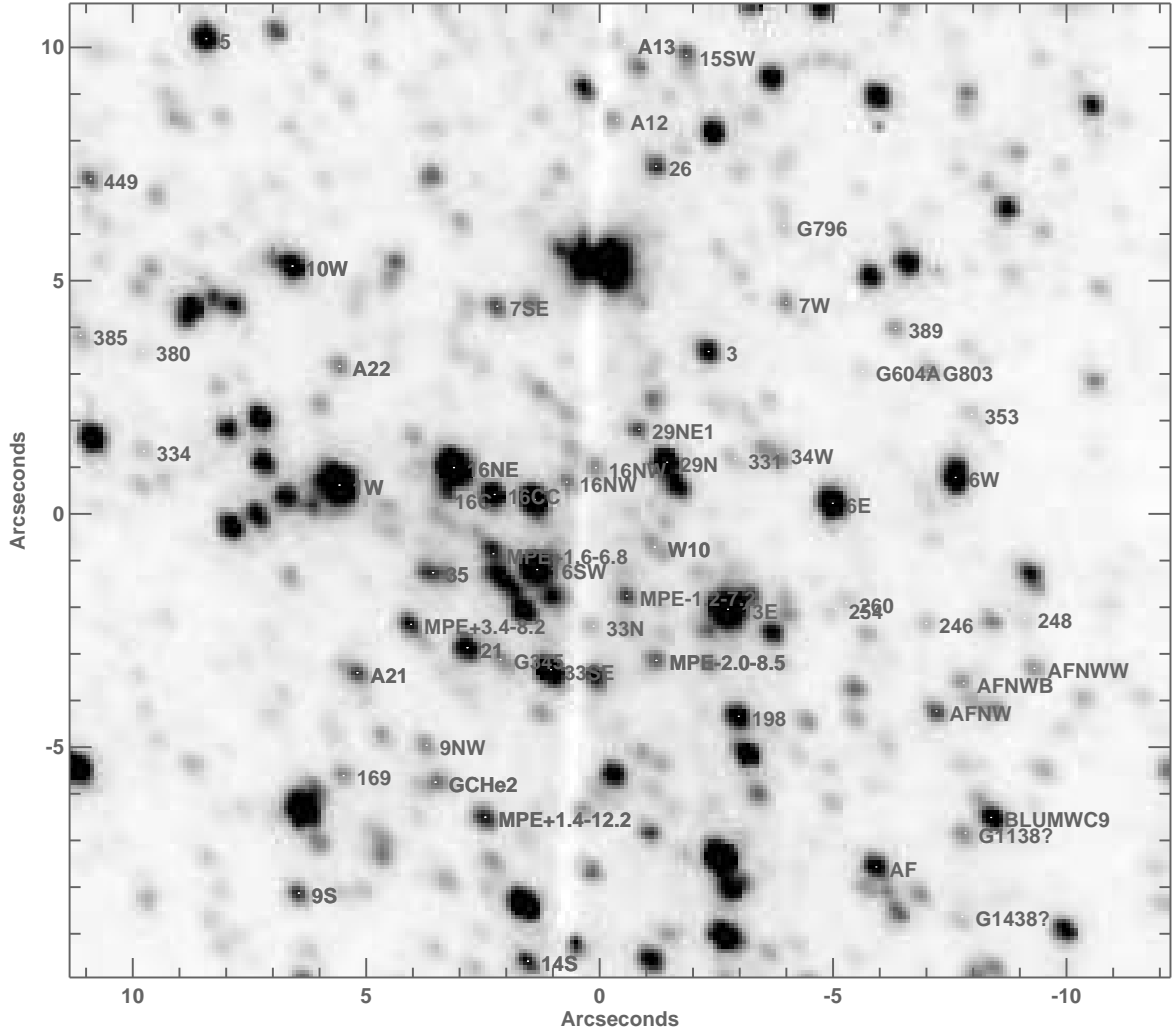}
\caption{Image taken with the
SCAM camera on NIRSPEC annotated with the names of those sources
that have been identified as massive stars in the literature. Those stars with
no common name published in the literature are given numbers based on their
order in our data files. The axes are
centered on the position of Sgr A*. North is up and East is to the left. The 
white strip through the middle of the image is the shadow of the slit. The
image has been rebinned by two for publication.
\label{image}}
\end{figure}

The observations were obtained on June 4, 1999, and July 4, 1999 with NIRSPEC, the
facility near-infrared spectrograph, on the Keck II telescope (McLean et al. 1998), in 
high-resolution mode covering K-band wavelengths (1.98 $\micron$ to 2.32 $\micron$).
The long slit (24'') was used in a north-south orientation, and the telescope was offset
by a fraction of a slit width to the west between exposures (see Figure~\ref{image}). 
Two data cubes were built from the spectra
obtained on the two nights. The field observed in July is almost entirely contained 
within the field
covered in June which covers a 24$\times$24 arcsecond region. 
Spectra of 40 stars, half of them hot, appear in both data cubes.

The slit viewing camera (SCAM) was used to obtain images
simultaneously with the spectra. These images make it easy to determine the slit orientation on
the sky when the spectra were obtained. 
From the SCAM images, we estimate seeing (FWHM) of 0$\farcs$6 on the first
night and 0$\farcs$4 on the second night. The plate scales for the spectrograph 
(0$\farcs$14/pixel along the spectral axis and 0$\farcs$2/pixel along the spatial axis) and imager (0$\farcs$18/pixel)
were taken from Figer et al. (2000). We chose to use the 5-pixel-wide slit (0$\farcs$72) for the
June slit scan and 3-pixel-wide slit (0$\farcs$43) for the July slit scan in order to match
the seeing on the respective nights. The corresponding resolving power was R$\sim$14,000 in June,
and R$\sim$23,300 in July, measured from sky OH and arc lamp lines of known wavelengths
in the two data sets. The slit positions were nearly parallel in the June scan, aligned
with the long axis along the north-south direction, while they were quite skewed in the
latter part  of the July scan because of problems with telescope tracking and/or image
rotator control during these commissioning runs. The NIRSPEC cross-disperser and the
NIRSPEC-7 and NIRSPEC-6 filters were used to image six echelle orders onto the 1024$^2$-pixel
InSb detector field of view. Note that the two filters were used in concert in July
in order to ensure complete rejection of light in unwanted cross-disperser orders.

\begin{figure}[ht]
\epsscale{0.95}
\plotone{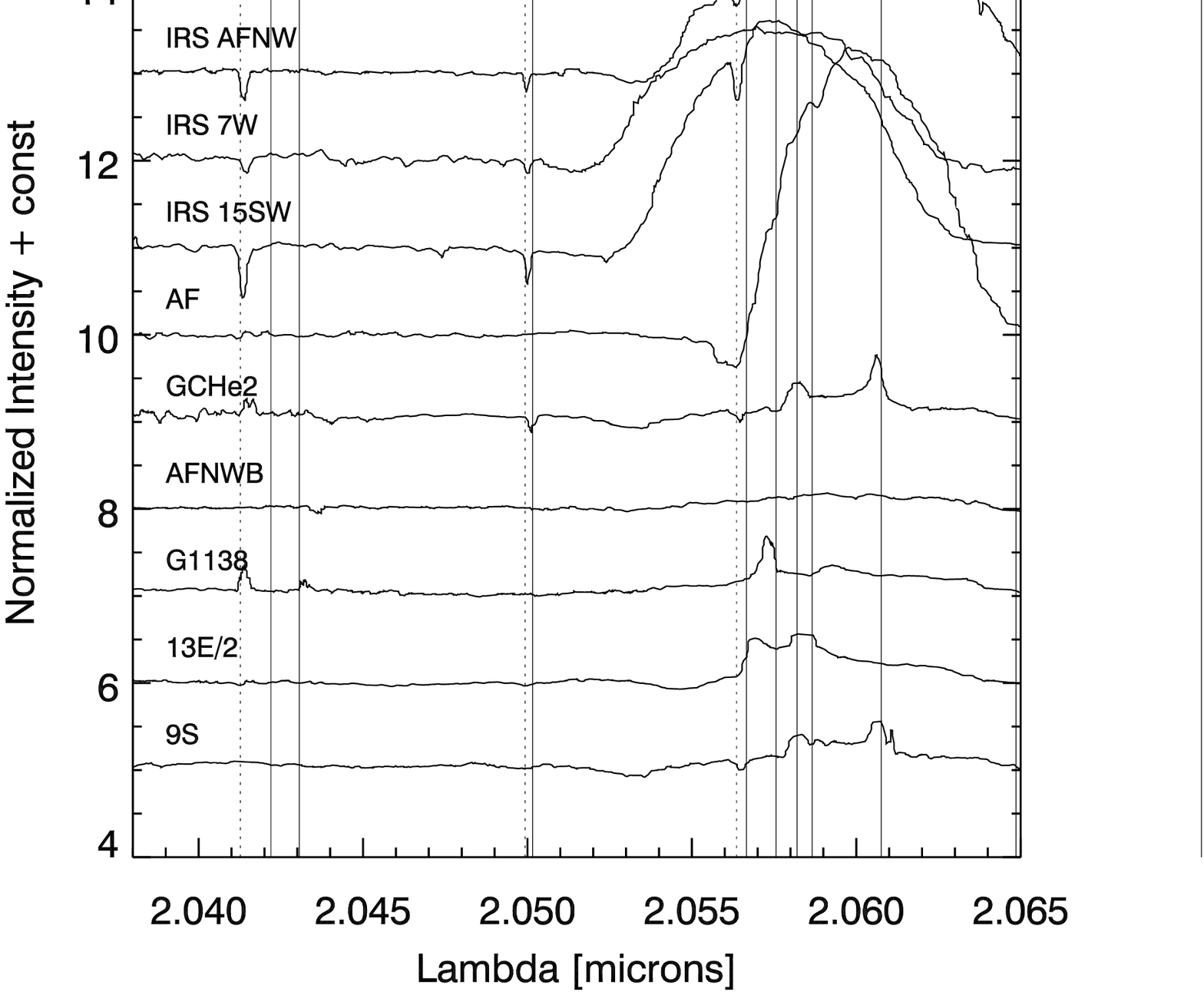}
\caption{Spectra of those
sources which have a broadened \ion{He}{1} 2.058 $\micron$ emission
line feature.  Also shown are the positions of the telluric features
which are not fully removed from the 
spectra (vertical dotted lines) and the stellar photospheric features
(vertical solid lines). There are also some narrow emission features
from the local ISM. The spectrum of IRS 13E has been divided by two
to fit on this scale.  \label{broadspec}}
\end{figure}

Some prominent features within the spectral range of our data set include diffuse and stellar lines of: 
Br $\gamma$
emission at 2.1661 $\micron$, \ion{Fe}{3} at 2.2178 $\micron$,
\ion{He}{1} at 2.0581 $\micron$, and the edge of the CO(2-1) transition at 2.2935 $\micron$.
Table~\ref{linetab} provides a list of known photospheric, telluric and interstellar medium 
(ISM) lines observed within the
wavelength ranges considered in this paper.
Quintuplet Star 3, which is featureless in this spectral region (Figer et al. 1998),
was observed as a telluric standard (Moneti et al. 1994). Arc lamps containing Ar, Ne, Kr, and Xe,
and spectra of the sky, were observed to set the wavelength scale. In addition, a continuum lamp was
observed through an etalon filter in order to produce an accurate wavelength
scale in between arc lamp and sky lines (predominantly from OH). A field relatively
devoid of stars, the ``dark spot" ($\alpha$=17h44m49.s, $\delta$=-28.54'6.''8, J2000),
was observed to provide a dark current plus bias plus background image.
A quartz-tungsten-halogen (QTH) lamp was observed to provide a flattened image.
There are still some visible telluric lines best seen in Figure~\ref{broadspec}
(shown with a vertical dotted line)
which did not go away completely from the sky calibration.

\section{DATA REDUCTION AND ANALYSIS}

We provide a brief overview of the data reduction procedure here with a
more complete discussion provided in Figer et al. (2003).
We removed sky emission, dark current, and residual bias, by subtracting images taken
at the location of the "dark spot". 
Due to the changing sky levels during the scan, the sky level in the dark sky image was
scaled before subtracting it from target images. The scale factor was chosen such that it
produced the minimum variance of residual sky line features in selected regions of the
image after flat-fielding. The images were then flattened by dividing
them by the QTH lamp image.
These flattened images were then cleaned of bad pixels using a two-pass procedure. First, we replaced
any pixel with a value five times the average of its neighbors within a 5 by 5 pixel area by the median of the neighbors
in that area. Second, we replaced pixels with values that were higher or lower than the values of
both immediate neighbors in the dispersion direction, and deviated by more than 10 times the
poisson noise from those two neighbors.

\begin{figure}[ht]
\epsscale{0.95}
\plotone{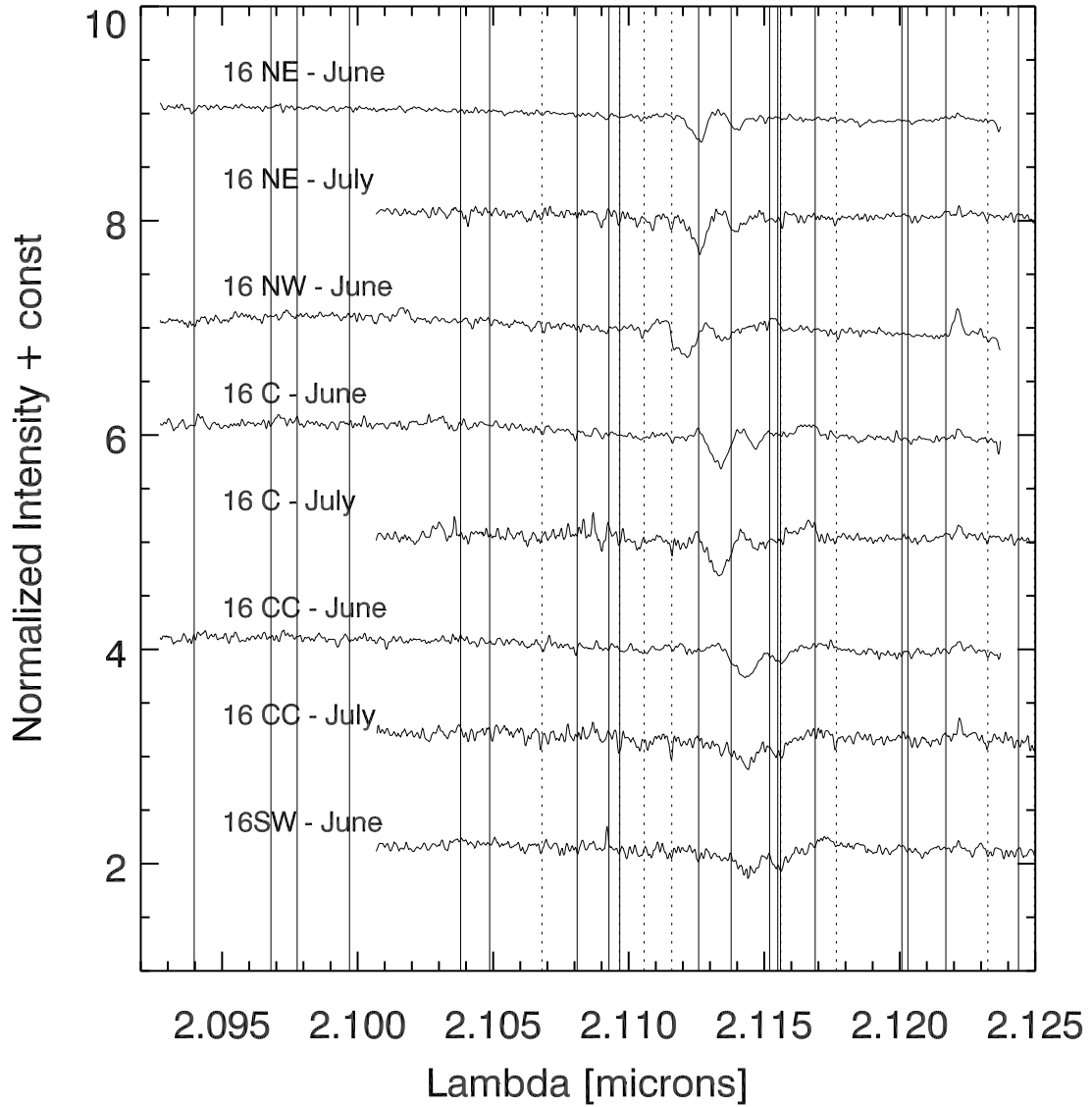}
\caption{Spectra of the IRS 16
sources which clearly show the \ion{He}{1} $\lambda\lambda$ 2.112,
2.113 doublet. Also shown in this and the remaining plots are the
positions of the telluric features (vertical dotted lines), some of which were not completely eliminated by the sky subtraction, and the
stellar photospheric features (vertical solid lines).
\label{narrowspec16}}
\end{figure}

\begin{figure}[ht]
\epsscale{0.95}
\plotone{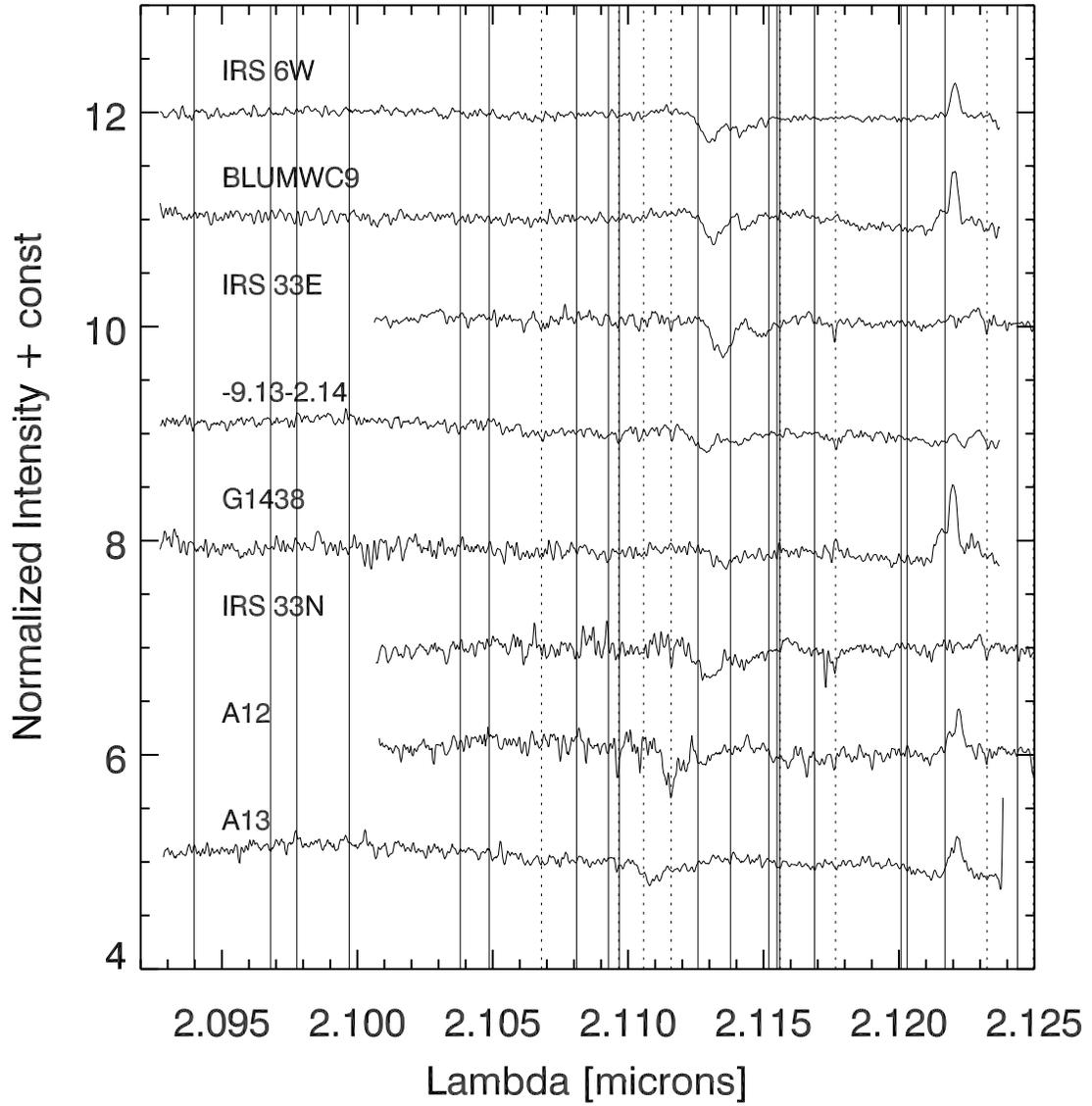}
\caption{Spectra of the
remaining sources which clearly show the \ion{He}{1}
$\lambda\lambda$ 2.112, 2.113 doublet. There are still some visible
telluric lines which did not go away completely from the sky
calibration. \label{narrowspec}}
\end{figure}

Out of the roughly 60 sources in our sample which have been classified as hot stars based on
previously published photometry and/or spectra, we identify
20 stars with spectral features falling into two distinct classes: 1) those
stars with \ion{He}{1} $\lambda\lambda$2.112, 2.113 $\micron$ doublet absorption
lines (see Table~\ref{nartab2} and Figures~\ref{narrowspec16} and~\ref{narrowspec}), and 2) those stars with
broad \ion{He}{1} 2.058 $\micron$ emission lines (see Table~\ref{brotab2} and Figure~\ref{broadspec}).
A list of all the notable stellar spectral features identified in this sample of stars is
given in Table~\ref{lineident}.

\begin{figure}[ht]
\epsscale{0.95}
\plotone{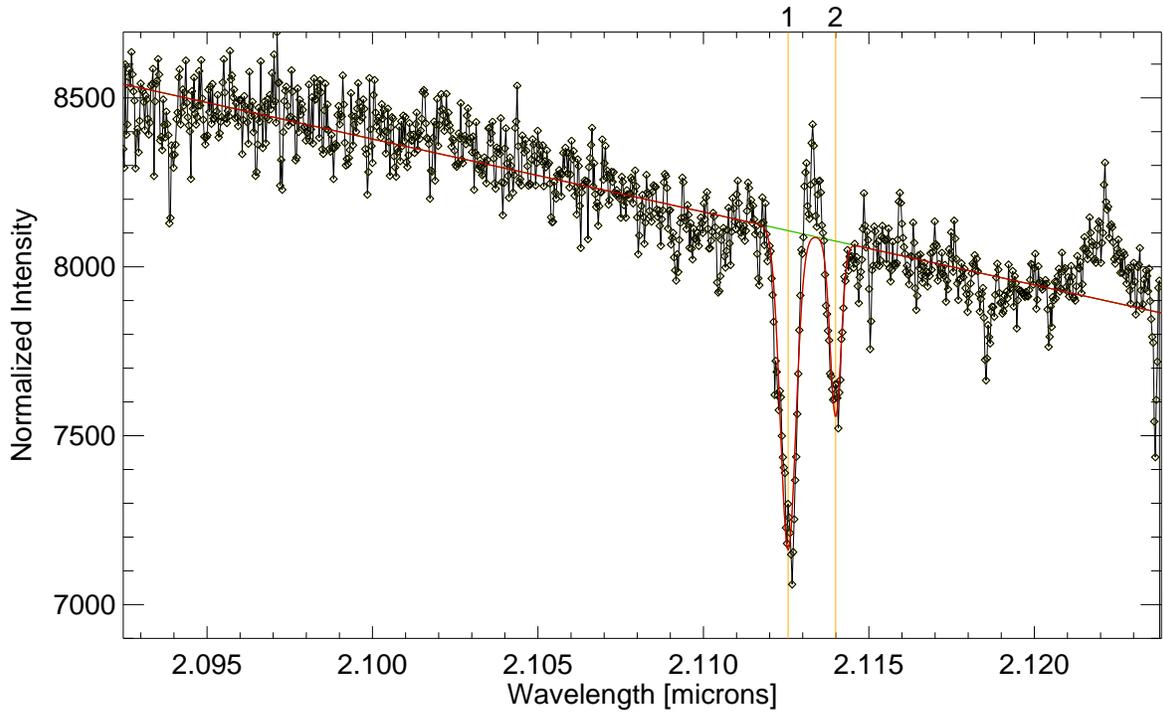}
\caption{The
\ion{He}{1} $\lambda\lambda$ 2.112 2.113 micron doublet feature
observed in the spectrum of IRS 16NE. The data are plotted with
diamonds, the continuum fit is the horizontally oriented dotted line and the gaussian
fits are the solid line. The centroids of the two gaussian fits are denoted
with the vertical dotted lines.
\label{narfitfig}}
\end{figure}

The radial velocities of the stars with the $\lambda\lambda$2.112, 2.113 $\micron$ doublet are initially 
estimated from the centroid of a gaussian fitted to each individual component of the 
doublet. First, the continuum level is modeled with
a first order polynomial and then the gaussian fits are applied to each of the absorption features 
using the IDL spectral line fitting program {\it lineplot}.
The two absorption lines produce two separate estimates for the radial velocity (see Figure~\ref{narfitfig}), and
when the doublet is detected in both the
June and July data sets, there are four separate estimates of the radial velocity.
The radial velocities estimated from the gaussian fits and their uncertainties are given in Table~\ref{nartab2}, 
which includes the average velocity estimated from the available measurements.
The uncertainties are calculated as either
the standard deviation of the values from four measurements (two lines, two observations) or
the half range of two measurements (two lines, one observation). The dominant source
of uncertainty is measurement error when fitting the line. If there were more lines
detected, then a cross-correlation of the observed spectrum with a model template would produce the
smaller uncertainties reported for the radial velocities of the cool stars from this 
same data set (Figer et al. 2003).
Most of the radial velocities of the He I emission-line stars estimated from the 2.112 and 2.113 $\micron$ lines
(with the exception of IRS 16C) are consistent with published values (P2001, see Table~\ref{nartab2}).
The radial velocities and classifications of these sources are more robust than those
of P2001 and G2000 since they are based on an
absorption feature, which is not as biased from local ISM emission as the 2.058 $\micron$ He I emission-line used
previously. 

\begin{figure}[ht]
\epsscale{0.75}
\plotone{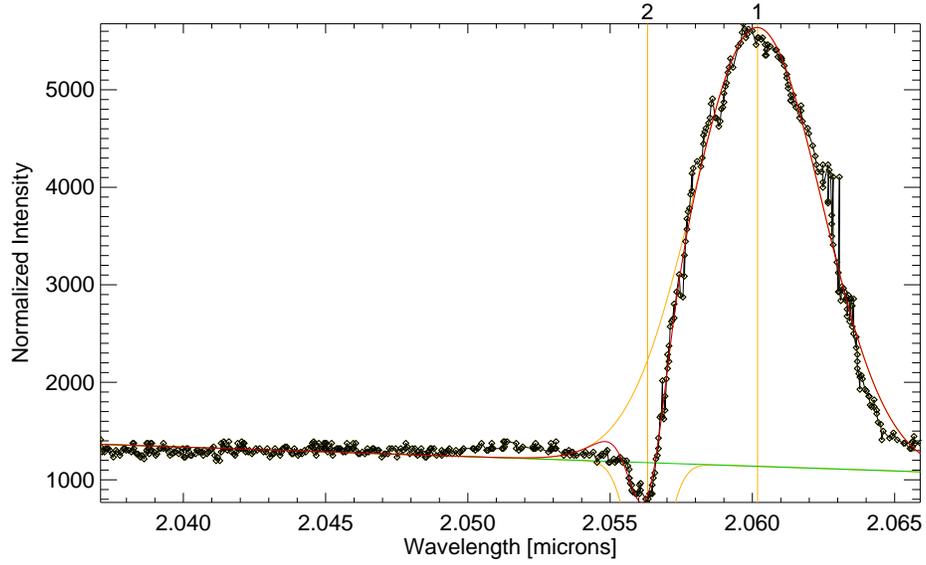}
\figcaption{The broadened He 2.058
$\micron$ emission line observed in the spectrum of the AF star. The
data are plotted with diamonds and the fit with a gaussian in
absorption on the blue side and a gaussian in emission on the red
side is shown by the solid line. The continuum fit is the horizontal dotted
line. The centroids of the two gaussian fits to the absorption and
emission features are denoted with the vertical dotted lines. \label{brofitfig}}
\end{figure}

\begin{figure}[ht]
\epsscale{1.0}
\plotone{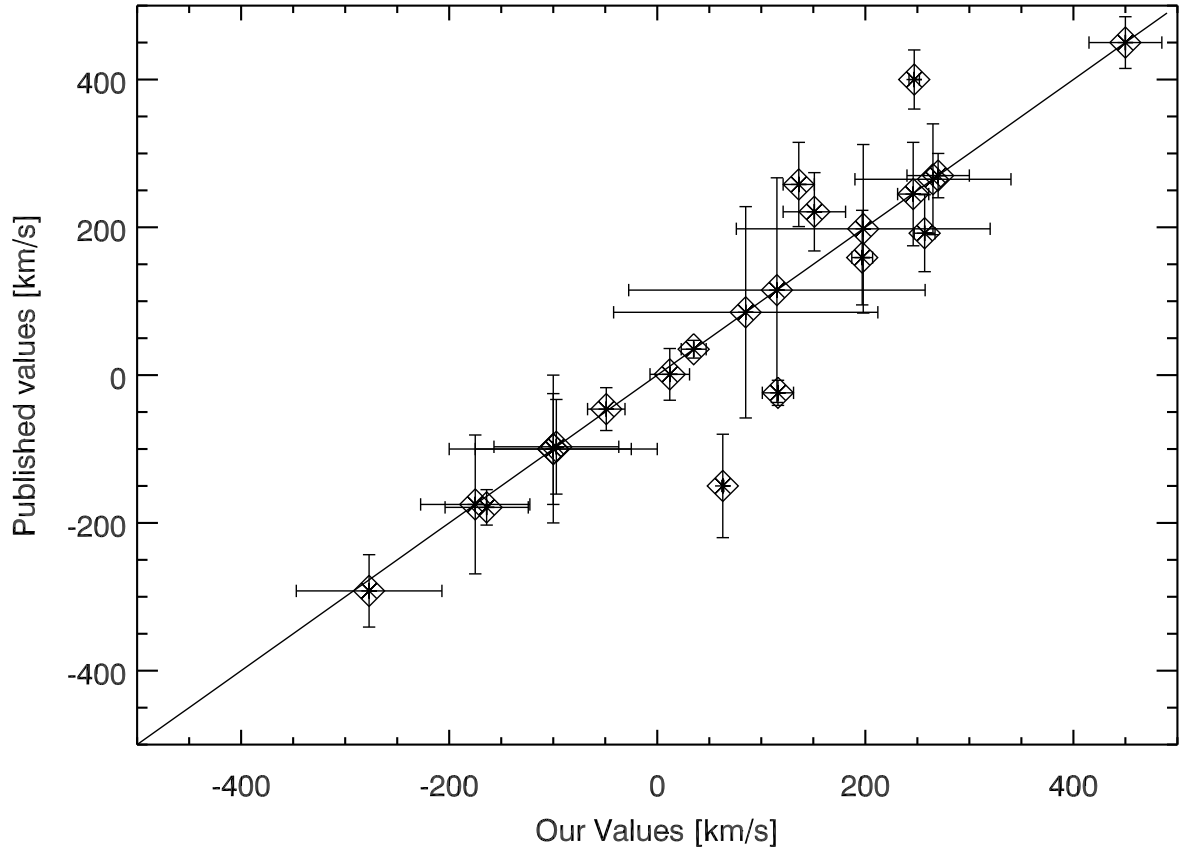}
\caption{Plot comparing our
derived radial velocity values to those from Paumard et al. (2001,
2003) or Genzel et al. (2000) showing good agreement between our
values and those published previously. \label{ourstheirs}}
\end{figure}

The broad \ion{He}{1} emission feature at 2.058 $\micron$ (see Figure~\ref{brofitfig}) 
is modeled as a single gaussian
component along with an additional gaussian absorption feature on the
blue side of the emission feature (a P Cygni profile).
The radial velocities estimated from the center of the emission feature are
listed in Table~\ref{brotab2} along with the FWHM of the emission feature thought to approximate the
outflow velocity of the stellar wind. For two sources (IRS 9S and G1138), 
the radial velocity was estimated from the
2.1125 $\micron$ \ion{He}{1} emission line due to low signal to noise in the 2.058 $\micron$ line. 
Since none of the broad-line sources are present in the July data set, 
we also estimate the radial velocities from the emission-line sources by cross-correlating the spectrum
around the 2.058 $\micron$ \ion{He}{1} emission line with a template consisting of a gaussian 
at the rest wavelength of the line and
the observed FWHM. The radial velocities estimated from both methods are
given in Table~\ref{brotab2} as well as their average and the half range.
These radial velocity measurements are also consistent with previously published values (P2001; G2000). 
Figure~\ref{ourstheirs} compares all the radial velocities estimated in this paper to their 
previously published values. We have adjusted all the radial velocities from the geocentric to the heliocentric frame by removing Earth motion,
+6.6 \kms  in June and -8.1 \kms  in July, where a positive value indicates motion toward the
Galactic center. The error in the resultant velocity due to this adjustment is estimated to be ±0.10 \kms.

\begin{figure}[ht]
\epsscale{1.0}
\plottwo{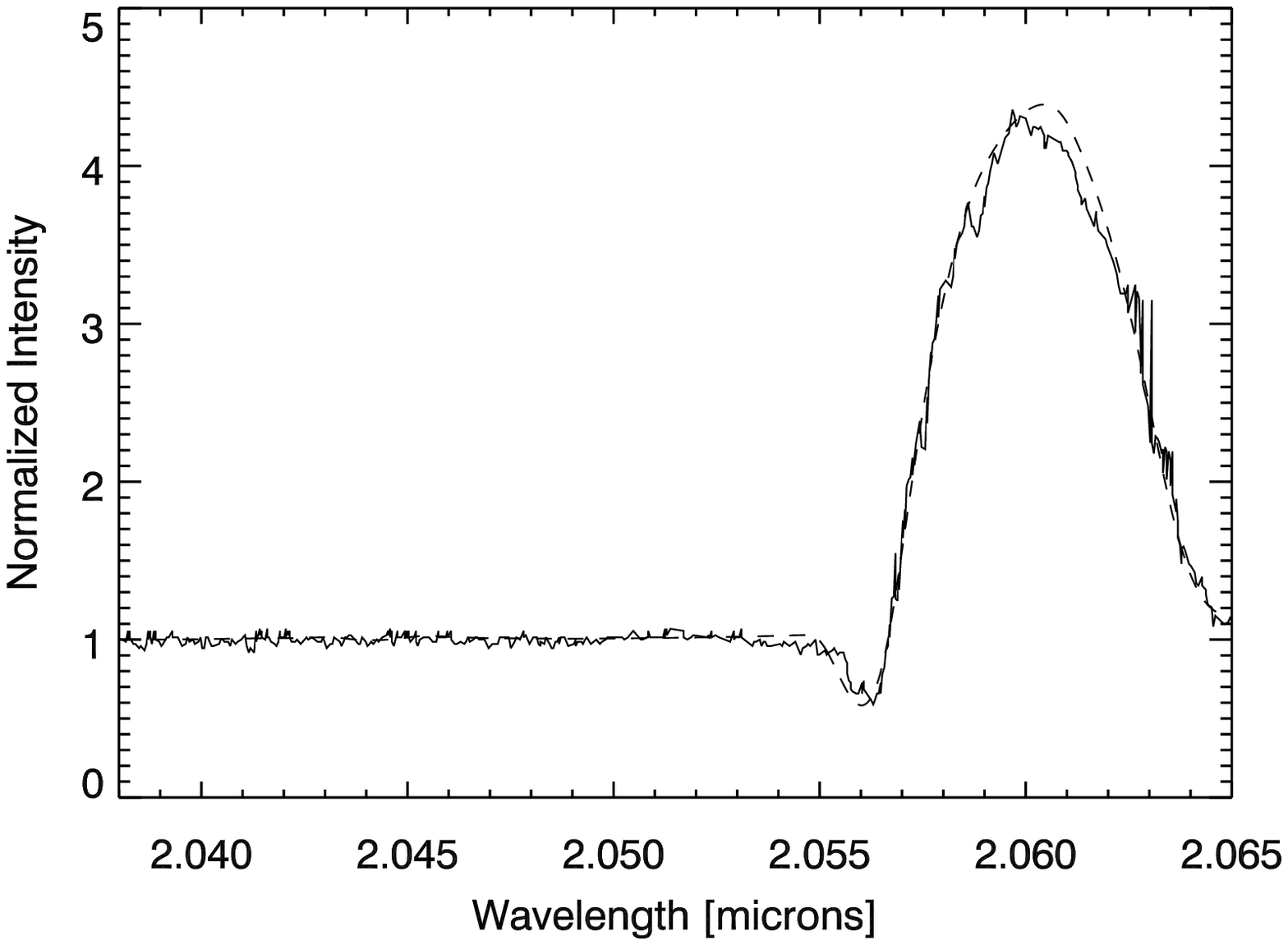}{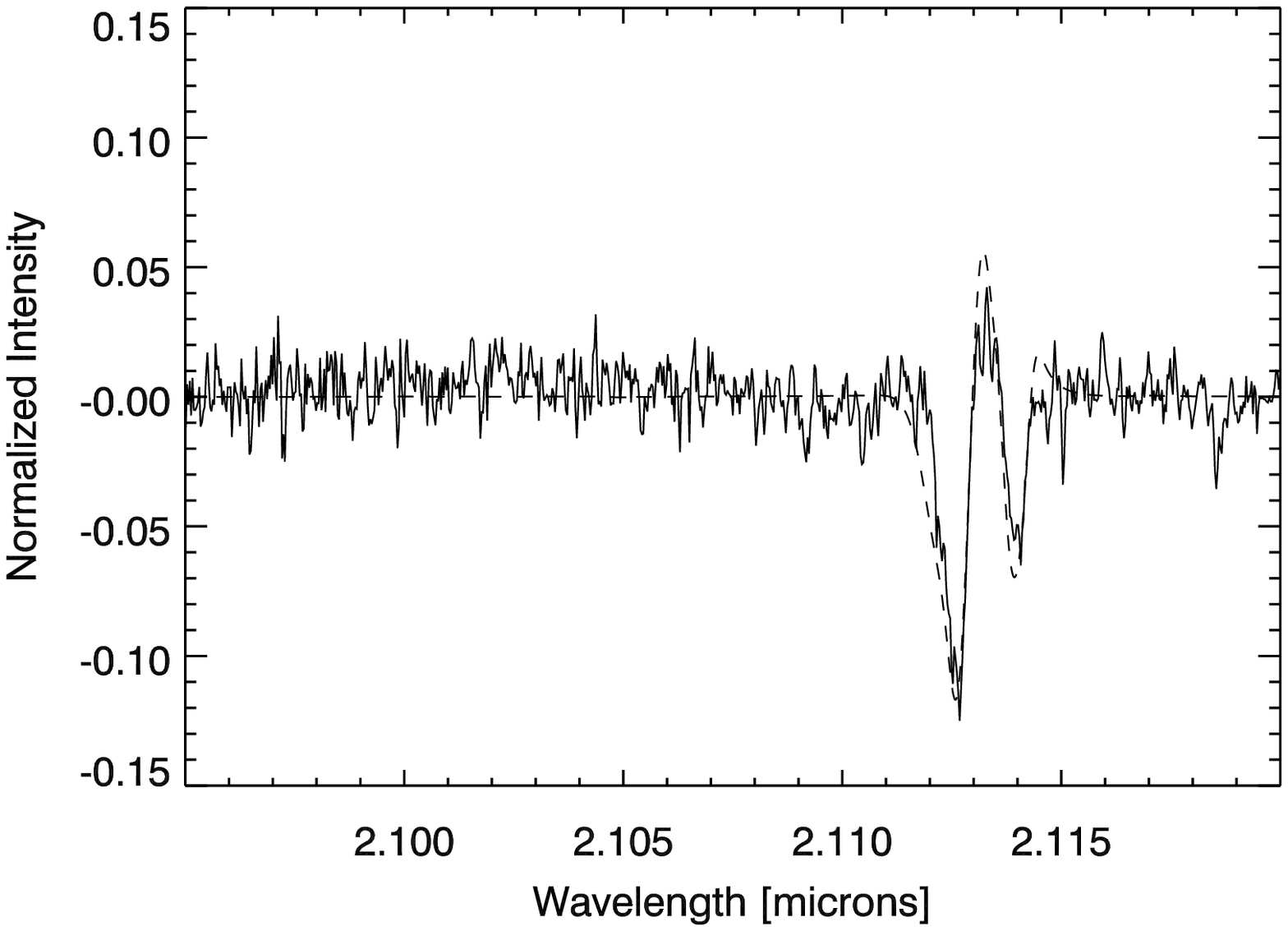}
\caption{Plots of the spectra of the AF star (left) and IRS 16NE (right) along with their
model fits (dashed line). The models have been shifted to the frame of the observed radial
velocities for each source. \label{modelanddata}}
\end{figure}

To verify that our line fitting procedure provides a valid radial velocity for each star, 
we also cross-correlated the spectra 
of IRS 16NE and the AF star with a non-LTE stellar atmospheric
model of the spectrum (Najarro et al. 1999, see Figure~\ref{modelanddata}). 
The model accounts for any effect the winds have on
the position and shape of the photospheric lines, such as blue shifting
the location of the minima of the absorption doublet. 
Also, when a P-Cygni profile is present, the radial velocity estimate from the theory always 
lies on the bluer side of the emission component and not at the position of the peak of the feature. 
Furthermore, the FWHM estimate from the broad emission line component of the P-Cygni 
profile is not necessarily a direct measure of the
terminal velocity of the wind. The case of the \ion{He}{1} 2.058 $\micron$ line is even worse since the absorption and
emission components will depend on wind parameters such
as mass-loss rate, the terminal velocity, and the shape of the velocity field.
From the cross-correlations, we estimate radial velocities of 49 and 239 \kms for
IRS 16NE and the AF star, respectively. This is an offset of 24 and 18 \kms from the values 
reported in Tables~\ref{nartab2} and \ref{brotab2}
estimated from gaussian fitting and gaussian cross-correlation. These offsets represent a potential 
correction factor to all the measured radial velocities due to the wind effects on the shapes of the spectral lines.  

\section{RESULTS}

In total we have added five newly classified He I emission-line stars (-9.13-2.14, A12, A13, G1438 and G1138) to the population
of such stars observed within the central few parsecs of the
Galactic center. This increases the total number of stars definitively identified as \ion{He}{1}
emission-line stars to around sixty when combined with
similar surveys, the most recent including Paumard et al. (2001, 2003, 2004), 
Genzel et al. (2000, 2003), and Ott (2004). The other two stars with new radial velocity
estimates are known He I emission line stars (AFNWB and GCHe2, Tamblyn et al. 1996).
With the aid of adaptive optics and/or larger telescopes, these high spatial resolution surveys have 
confirmed the identities of many well known He emission-line stars classified originally based on lower
spatial resolution observations 
(Krabbe et al. 1995; Tamblyn et al. 1996; Libonate et al. 1995; Najarro et al. 1997
and references within).
The Paumard et al. surveys covered
an area of 40$\times$24 arcseconds with 38 He I emission-line star detections 
separated into two primary categories - broad and narrow emission-line stars
based on the FWHM of their \ion{He}{1} 2.058 $\micron$ emission line.
P2001 used both their spectral features and difference in luminosity (the broad-line
stars are fainter on average than the narrow-line stars) to classify the
narrow-line stars as LBV-like stars and the broad-line stars as Wolf-Rayet stars.
The survey by Ott (2004) shares many sources with the Paumard et al. survey
with 26 He I emission-line star detections. They used the three measured velocity components
to separate the stars into clockwise and counter-clockwise rotating
populations based on their normalized angular momenta. A recent high spatial resolution study of the stellar population
within the central parsec done with the VLT SPIFFI integral
field spectrograph has identified IRS 16SE2 as a WN 5/6 star (Horrobin et
al. 2004). This study also classified IRS 3 as a WC 5/6 star, however, its classification is
still under debate (Pott et al. 2005).
A list of those stars with {\it published} radial velocities which were not
observed in this work is given in Table~\ref{remtab}, providing a complete list of radial velocities
for the identified \ion{He}{1} emission-line stars in the central parsec. For those stars
with no available spectra, Ott (2004) also identifies $\sim$400 stars
as potential He I emission-line stars or main-sequence, early-type stars based on a CO-band index estimated
from 2.29 and 2.26 $\micron$ AO images available from the
Gemini North Galactic center Demonstration Science data set.
Finally, Tanner et al. (2005) added six stars embedded within the Northern Arm to
the population of early-type stars using the standoff distances
and position angles of the bowshock surrounding these sources.

\begin{figure}[ht]
\epsscale{0.95}
\plotone{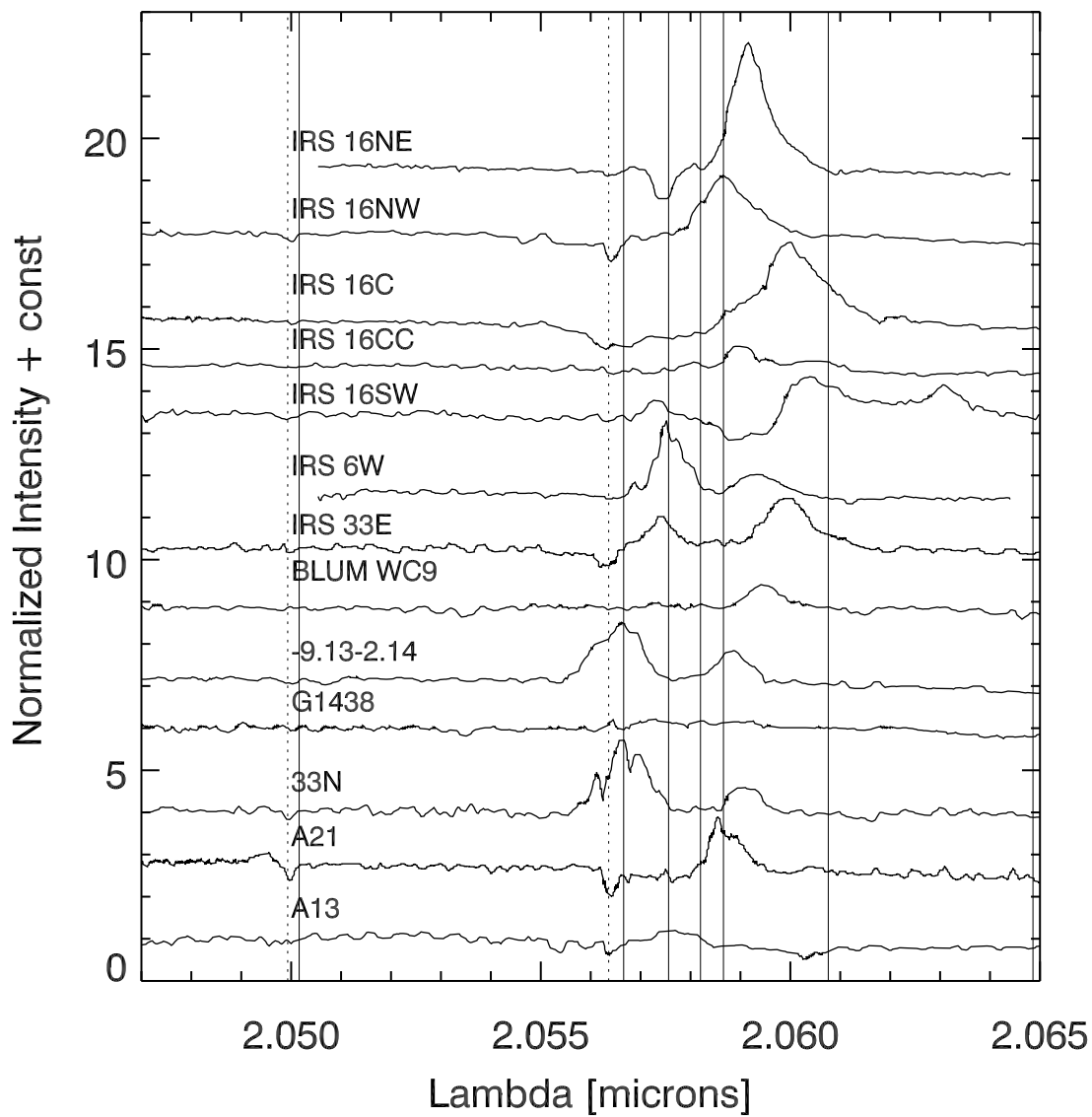}
\caption{Spectra of those sources which have the \ion{He}{1}
$\lambda\lambda$ 2.112, 2.113 doublet at the locations of the \ion{He}{1} 2.058 $\micron$ emission
line feature. The emission line features seen here are due to gas emission from
the mini-spiral which permeates throughout the region. These spectra were smoothed by
a factor of two and the uncorrected sky lines were removed. 
Also shown are the positions of the telluric features which were not
fully removed from the spectra
(vertical dotted lines) and the stellar photospheric features
(vertical solid lines).  \label{narrowspec2}}
\end{figure}

\begin{figure}[ht]
\epsscale{0.95}
\plotone{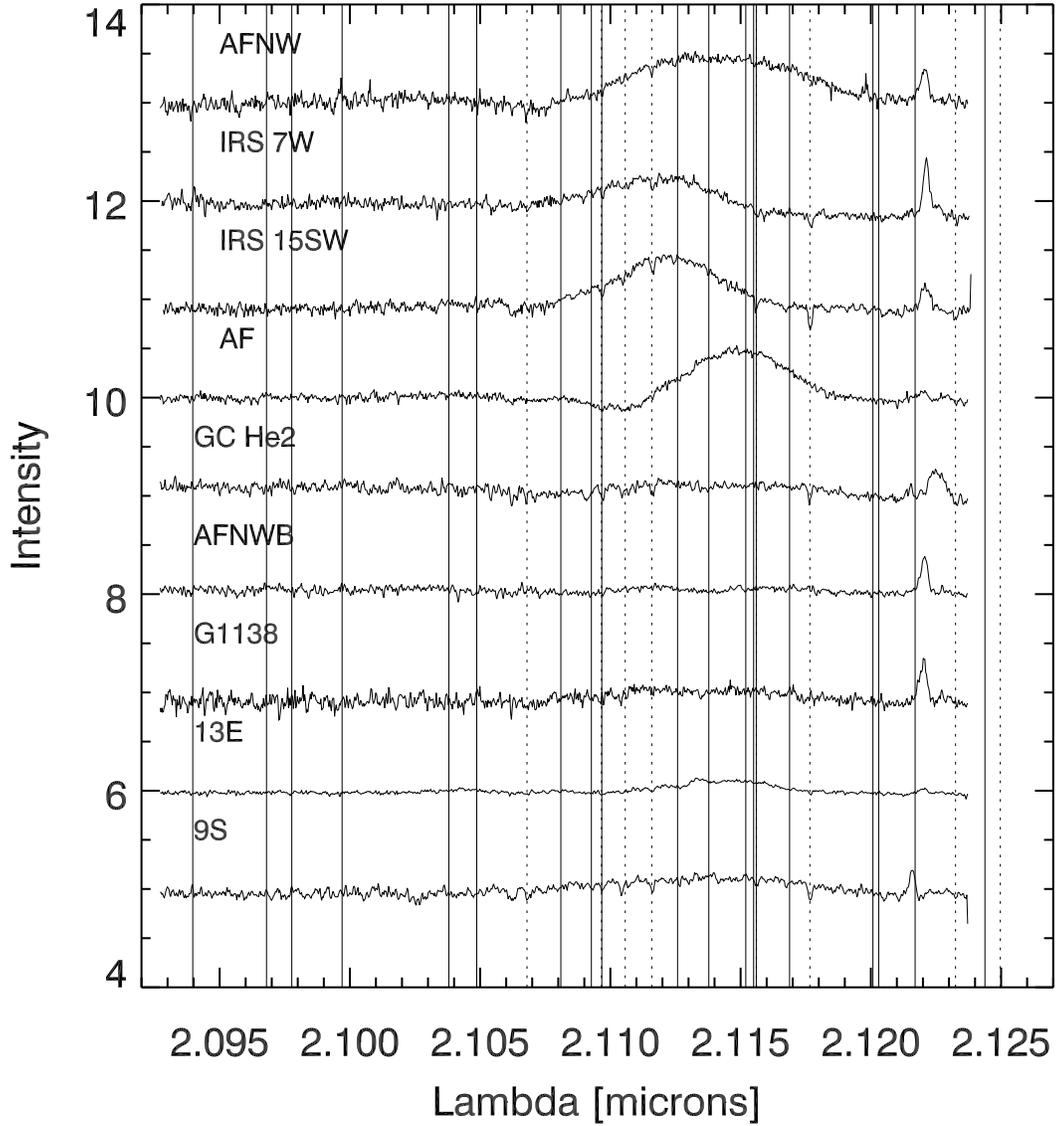}
\caption{Spectra of those
sources which have a broadened \ion{He}{1} 2.058 $\micron$ emission
at the location of the \ion{He}{1} $\lambda\lambda$ 2.112, 2.113 line.  
Also shown are the positions of the telluric features
(vertical dotted lines) and the stellar photospheric features
(vertical solid lines).  \label{broadspec2}}
\end{figure}

The types of spectral features observed around these stars fall into two distinct classes: 
1) those stars with the \ion{He}{1} $\lambda\lambda$ 2.112, 2.113 $\micron$ absorption doublet, and 2) those with
the \ion{He}{1} 2.058 $\micron$ broad emission line. 
For comparison, Figures~\ref{narrowspec2} and~\ref{broadspec2} show the spectra of the stars with
either the \ion{He}{1} $\lambda\lambda$ 2.112, 2.113 $\micron$ doublet 
or the \ion{He}{1} 2.058 $\micron$ broad emission line at the 2.058 and 2.112/2.113 spectral
region, respectively, to show that the stars with one
of the two designations do not have the other type of spectral feature. 
Most of the sources in both classes also exhibit complex Br $\gamma$ emission line features at around 
2.058 $\micron$ due to emission
from the diffuse gas within the mini-spiral which pervades the region. 
IRS 16NE also has some additional \ion{He}{1} absorption features near the Br $\gamma$ line due to
hydrogenic helium. When we consider both the total the equivalent width of both lines in the
\ion{He}{1} $\lambda\lambda$ 2.112, 2.113 $\micron$ feature
(provided in Table~\ref{nartab1}) and the absence of any detectable 
\ion{N}{3} 2.1155 $\micron$ emission, we determine
that the stars with the \ion{He}{1} absorption doublet have the spectral features resembling late O to early B stars similar to the
stars within one arcsecond of Sgr A* (Hansen et al. 1996; Ghez et al. 2003; Sch\"{o}del et al. 2003).  

\begin{figure}[ht]
\epsscale{1.0}
\plotone{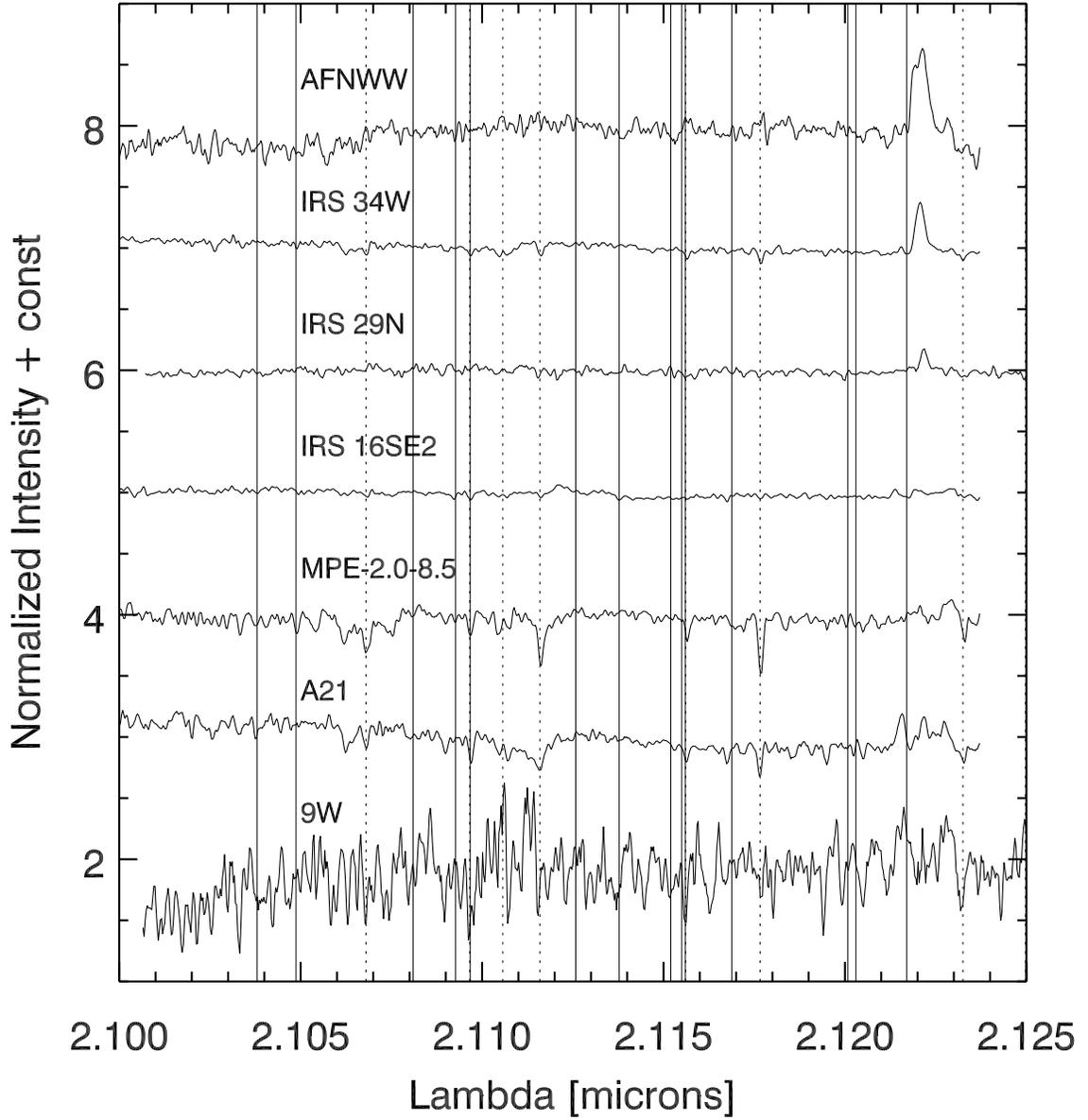}
\caption{ Spectra of
AFNWW, 29N, 16SE2 and 34W taken near the He I $\lambda\lambda$ 2.112,
2.113 $\micron$ \ion{He}{1} absorption doublet. Paumard et al.
(2003) classify IRS 34W as a narrow line star based on its 2.058
$\micron$ emission line, however, for this source we detect no 
He I $\lambda\lambda$2.112,2.113 $\micron$
absorption feature, even though it is seen in the remainder of
our spectra of Paumard's narrow-line sources. \label{extraspec2}}
\end{figure}

\begin{figure}[ht]
\epsscale{1.0}
\plotone{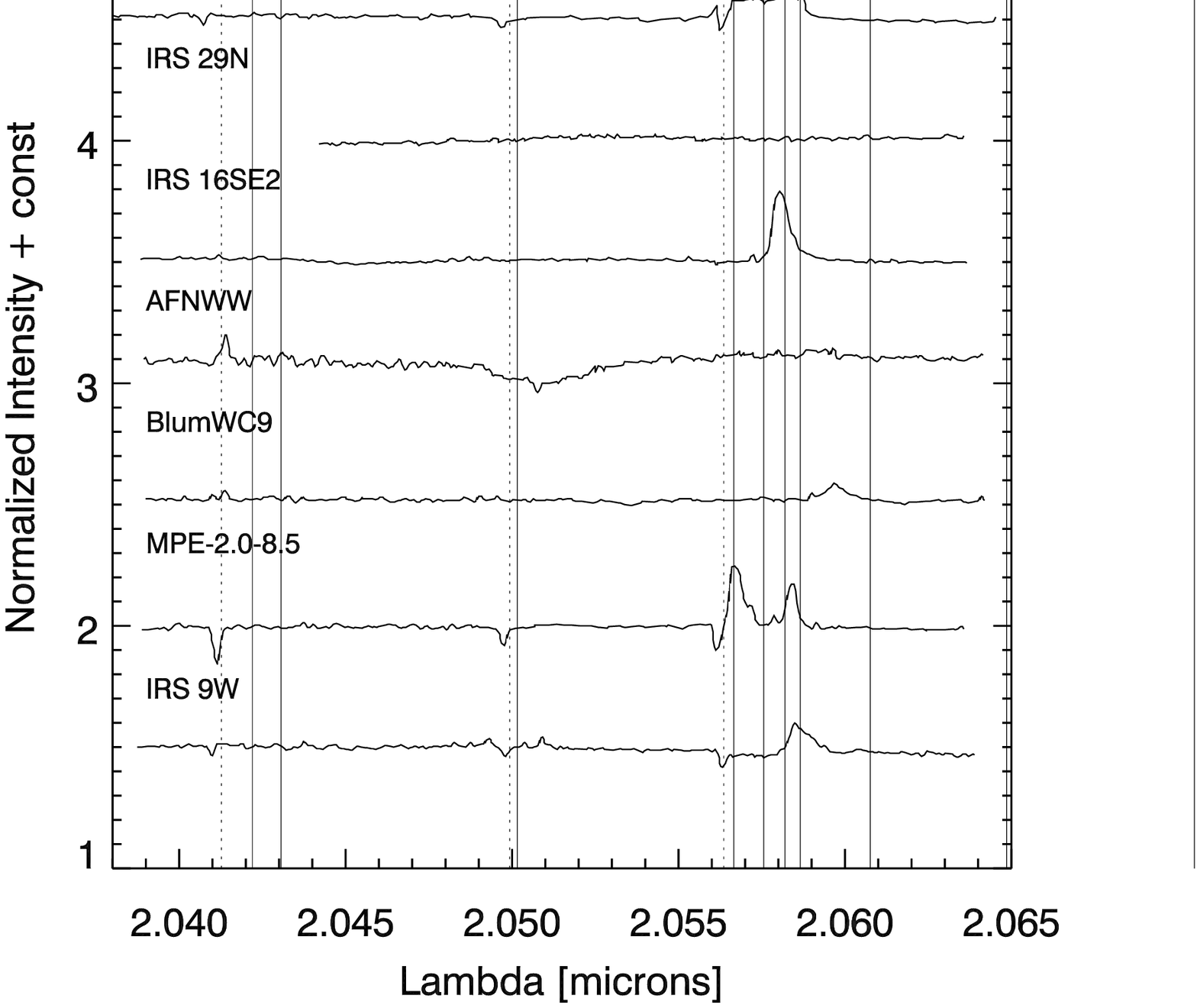}
\caption{ Spectra of
AFNWW, 29N, 16SE2 and 34W taken near the 2.058 $\micron$ \ion{He}{1}
emission line. The star, AFNWW, has a broad absorption feature near
2.058 $\micron$ but lacks the associated emission feature presumably
due to its wind properties. Many spectra also have narrow gas emission features originating
from the local ISM.  \label{extraspec1}}
\end{figure}

Some sources in our sample are notable for having no detectable emission or
absorption features falling in the two classes mentioned above or for
having published classifications which contradict our results.
IRS 34W has been classified as a narrow-line star based on
its 2.058 $\micron$ emission line and a variable star based on its light curve 
(Paumard et al. 2003;  Ott et al. 2003), however, 
we detect no $\lambda\lambda$ 2.112, 2.113 $\micron$ absorption
feature for this source (see Figure~\ref{extraspec2}), as is seen in our spectra of the other "narrow-line" sources classified in 
Paumard et al. (2001, 2003).
We do not observe a distinct broad \ion{He}{1} emission line in the spectra of either 
IRS 29N or 16SE2 other than some narrow-line emission created by the surrounding ISM (see Figure~\ref{extraspec1}). 
These two stars are
classified as broad-line sources in P2003 based on the FWHM of their 2.058 $\micron$ He I emission line.
Blum et al. (1995) originally classified the source BSD WC9, also known as Blum WC9, as a WC9 Wolf-Rayet 
star based on the detection of \ion{C}{4} and \ion{C}{3} lines with  
P2003 providing the more general classification of a broad-line He I star.
While we observe no broad He I emission line, we do detect the $\lambda\lambda$ 2.112, 2.113 $\micron$
absorption feature which suggests a potential misclassification for this star (see Figures~\ref{narrowspec} and~\ref{extraspec1}). 
One alternative to miss-classification for these sources is that, 
as a result of their variable nature,
their spectral features changed over the duration of the different observing epochs as has been
observed in stars of similar spectral type (Morris et al. 2004). 
Figure~\ref{extraspec1} shows a portion of the spectrum of AFNWW which has a 
distinctive absorption feature with no trace of the expected emission component.
Two of the stars with new radial velocity detections (G1138 and G1438) were
classified at "CO" stars in Genzel et al. (2000).
We have also observed the two stars which lie at the center of the mini-cavity
and Eastern Cavity (Stolovy et al. 1999). While both MPE-2.0+8.5 and A21 are definitely hot
stars based on a lack of CO absorption in their NIRSPEC spectra, we do not observe any stellar
photospheric features from which we would be able to estimate
a mass-loss rate. The question remains as to whether or not
these stars are the sole contributors to the winds necessary to
create the cavities in which they are embedded and whether the cavities exist at all (Paumard et al. 2004).
If  A21 and MPE-2.0+8.5 are truly hot stars, and are truly at the
centers of the mini-cavity and Eastern Cavity, respectively, then they might excite
those cavities. Finally, no distinguishable photospheric lines were detected in any of the
Northern Arm sources or IRS 3.

Table~\ref{nondet} presents a list of those stars which are classified as
hot stars based on their lack of prominent CO absorption features, 
but which do not have any distinguishable
photospheric lines in our data set. Those sources which do not have 
any previously published name were assigned a designation based on 
their RA and Dec offsets from Sgr A* in arcseconds.
Some of the stars with detectible He I emission lines also show a noticeable
dip in the region of the CO absorption features, but the weak strength of this feature and its
presence in definitively classified He I emission line stars suggests that the absorption is most likely the result of
pollution from the numerous background red giants comprising the older stellar population in the Galactic center (Haller et al. 1996).

\begin{figure}[ht]
\epsscale{1.0}
\plotone{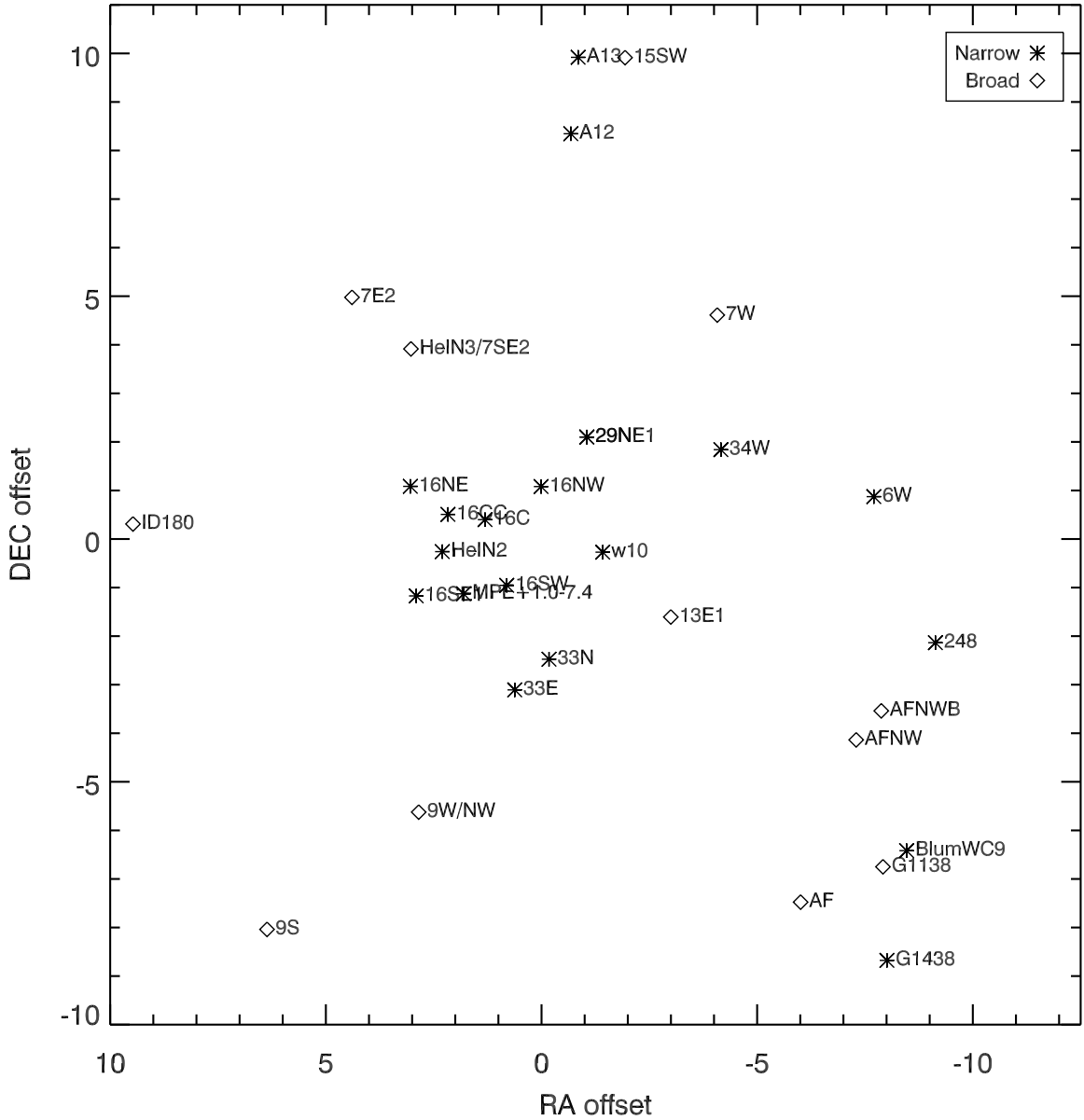}
\caption{Plot of the spatial
distribution of He emission-line stars detected to date. The broad emission-line stars
are denoted with an asterisk while the narrow emission-line stars are denoted with a diamond. 
The axes are centered on the position of Sgr A*. North is up and East is to the left.
\label{heoffset}}
\end{figure}

By increasing the sample of He I emission-line stars, we can look for
trends in the spatial distribution and kinematics of these stars. Figure~\ref{heoffset} shows
the spatial distribution of the sources detected in this paper, P2001 and G2000.
Paumard et al. (2001) claim that the narrow- and broad-line sources are distributed 
in two distinct centralized and ring-like populations on the plane of the sky over a
region similar to that covered in this study. 
The distinction in the spatial distribution of the narrow- and broad-line sources is not as discernable in 
Figure~\ref{heoffset}, with the addition
of a few narrow-line stars at distances greater than five arcseconds from Sgr A*.
However, the prominence of the IRS 16 cluster of narrow-line sources near
Sgr A* is still impressive and is not biased in our survey due to a drop off of sensitivity since
the slit was scanned over a large field of view. To quantify any true asymmetry in the
spatial distribution of the stars on the plane of the sky, we calculate
an asymmetry factor, Q, defined as the difference in the number of broad- and narrow-line stars within 
and beyond a set of specified radii. 

\begin{equation}
Q_i = (n_{narrow}-n_{broad})_{r<r_i} - (n_{narrow}-n_{broad})_{r>r_i}
\end{equation}

\begin{figure}[ht]
\epsscale{0.9}
\plottwo{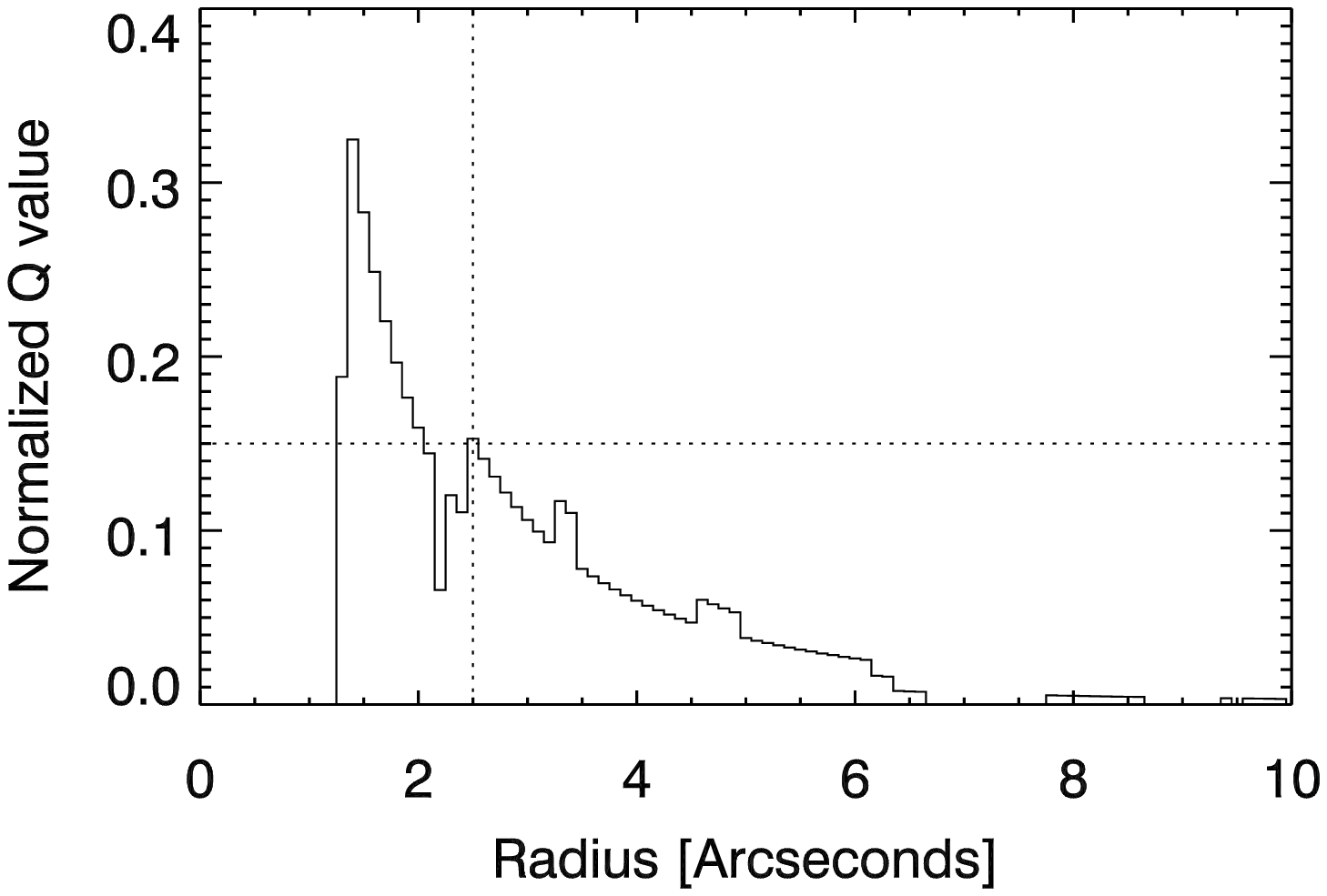}{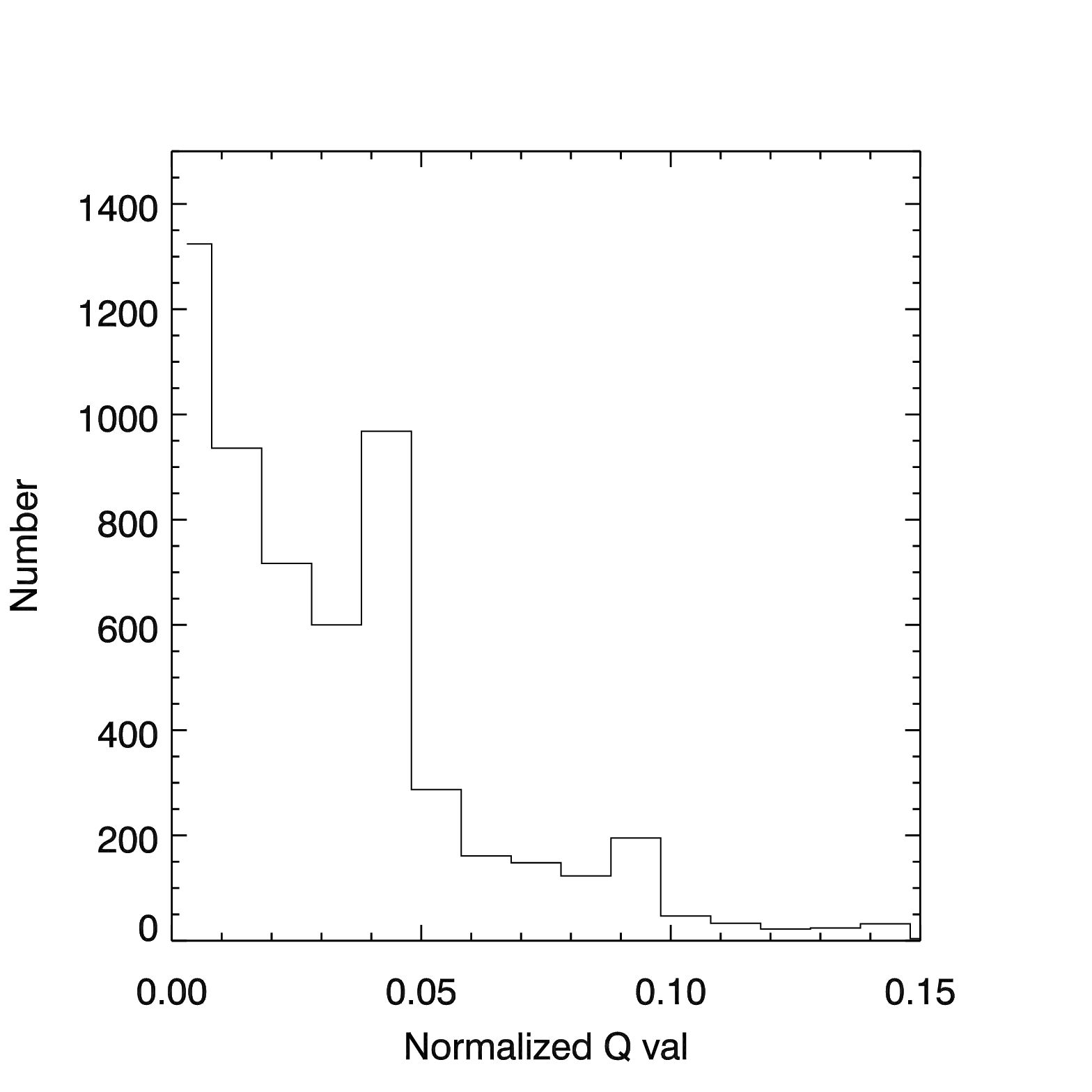}
\caption{{\it Left} - Plot of the values of the asymmetry factor, Q, estimated from our data as a function of 
radius from Sgr A*. The peak of Q=0.15 at 2.2 arcseconds is highlighted. 
{\it Right} - Histogram of the values of the same asymmetry factor estimated from 
10,000 Monte Carlo iterations. The asymmetry factor only reaches a value of 0.15 or more
less than 1\% of the time suggesting that the stars in the central few arcseconds
are not randomly distributed on the plane of the sky  \label{qplots}}
\end{figure}

The differences are normalized by the area of the inner and outer regions. 
The set of asymmetry factors estimated from the data as a function of distance from Sgr A* 
has a distinctive peak at a radius of about 2$\farcs$5 (see Figure~\ref{qplots}) after 
falling off from Sgr A*. When we run ten thousand iterations of a Monte Carlo code in which the stars 
are placed randomly within
the same field-of-view as the data, the same or greater value of the asymmetry factor as that derived from the
data only occurs 0.5\% of the time (see Figure~\ref{qplots}), suggesting that the distribution of 
stars within the central parsec is not a random distribution. 

\begin{figure}[ht]
\epsscale{0.9}
\plotone{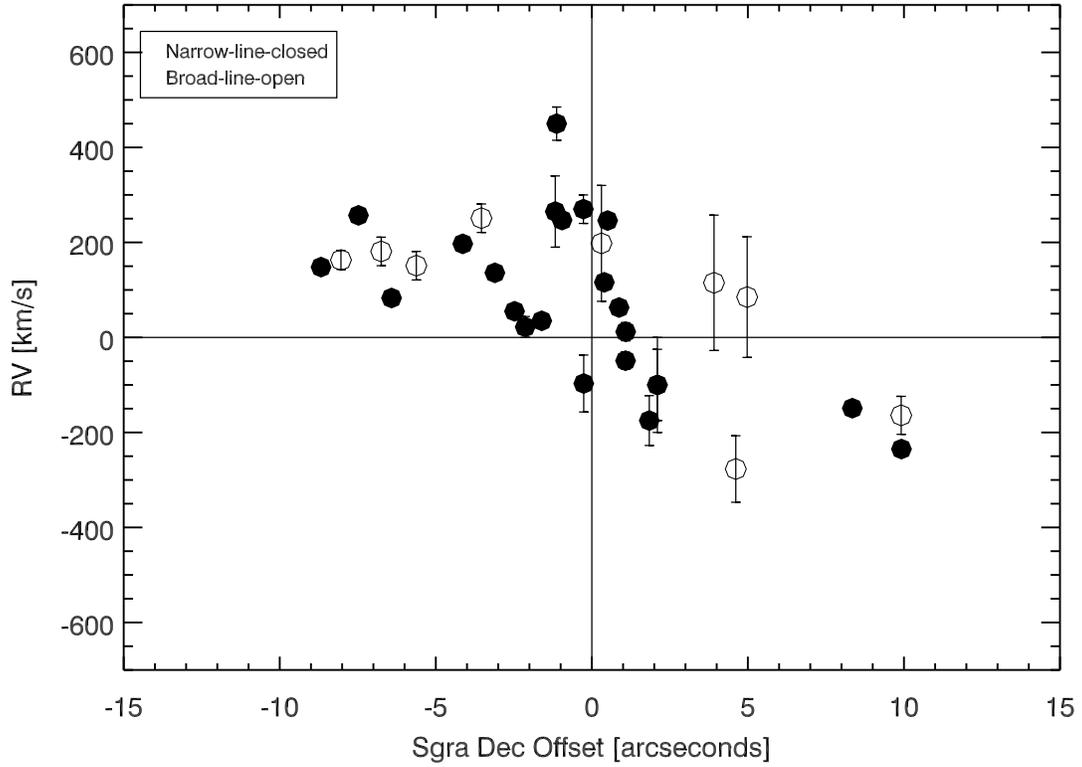}
\caption{ Plot of the radial
velocities of each star as a function of declination from Sgr A*. The narrow-line stars are plotted
with solid circles and the broad-line stars are plotted with open circles. There appears to be
a trend for the sources to lie in the upper left and lower right
panels, consistent with the suggestion that the stars are orbiting in a 
coherent disk (Genzel et al. 2000; Levine and Beloborodov 2003). \label{rvvsdec}}
\end{figure}

With the addition of seven new RV measurements, we can also compare the kinematics of the broad- and narrow-line populations using the stars with 
both proper motion and radial velocity measurements given in this paper and in the literature. 
Figure~\ref{rvvsdec} shows a plot of the measured radial velocities as a function of offset
from Sgr A*. This plot  includes both the narrow-line (filled circles) and broad-line
sources (open circles) with no significant difference in the velocity offsets for these two
populations. Most of the stars lie along a diagonal going through the position of Sgr A*, supporting previous
claims that these stars comprise at least one coherent rotating disk (Levin \& Beloborodov 2003; P2001; G2000).

\begin{figure}[ht]
\epsscale{1}
\plotone{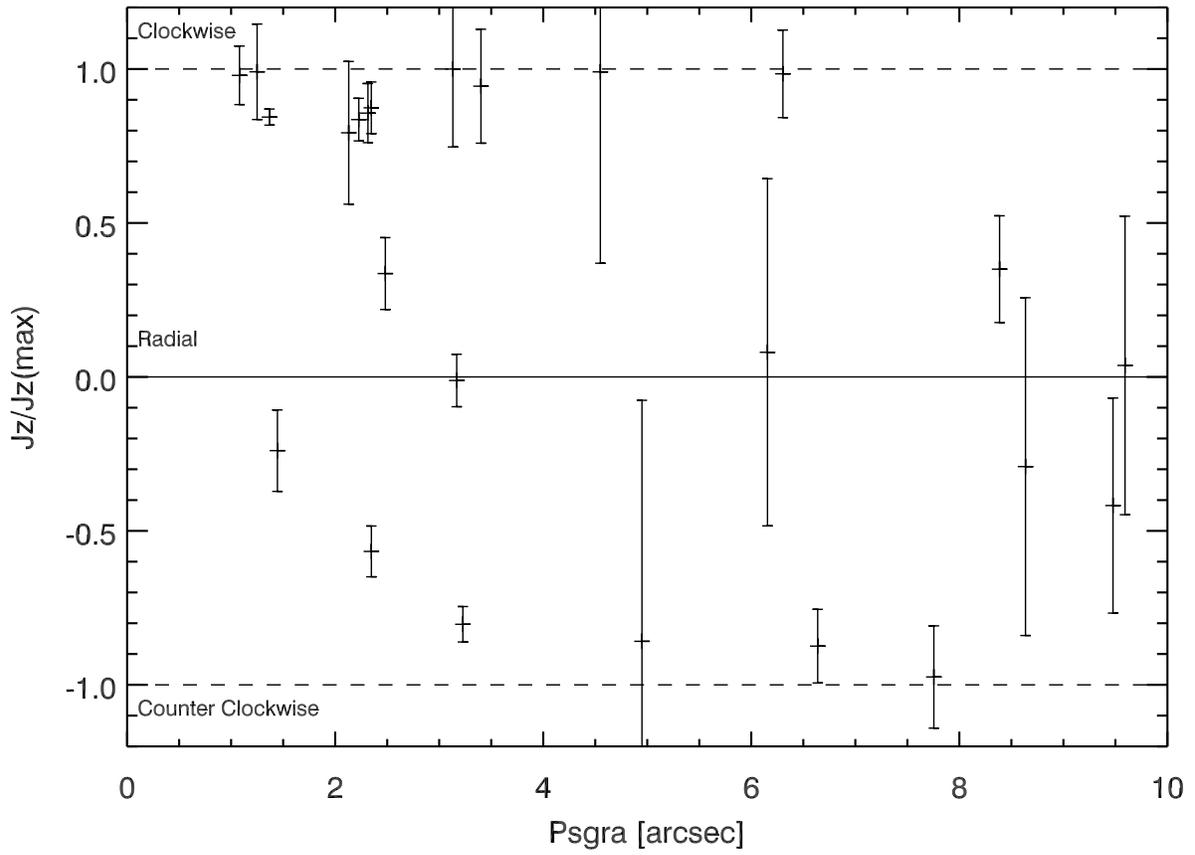}
\caption{ Plot of the normalized angular momentum as a function of distance
from Sgr A* in arcseconds. Those stars with J$_z$/J$_z(max)$ $>$ 0 are in clockwise orbits while
those with J$_z$/J$_{z(max)}$ $<$ 0 are in counter-clockwise orbits. \label{jmaxplot}}
\end{figure}

\begin{figure}[ht]
\epsscale{1.0}
\plottwo{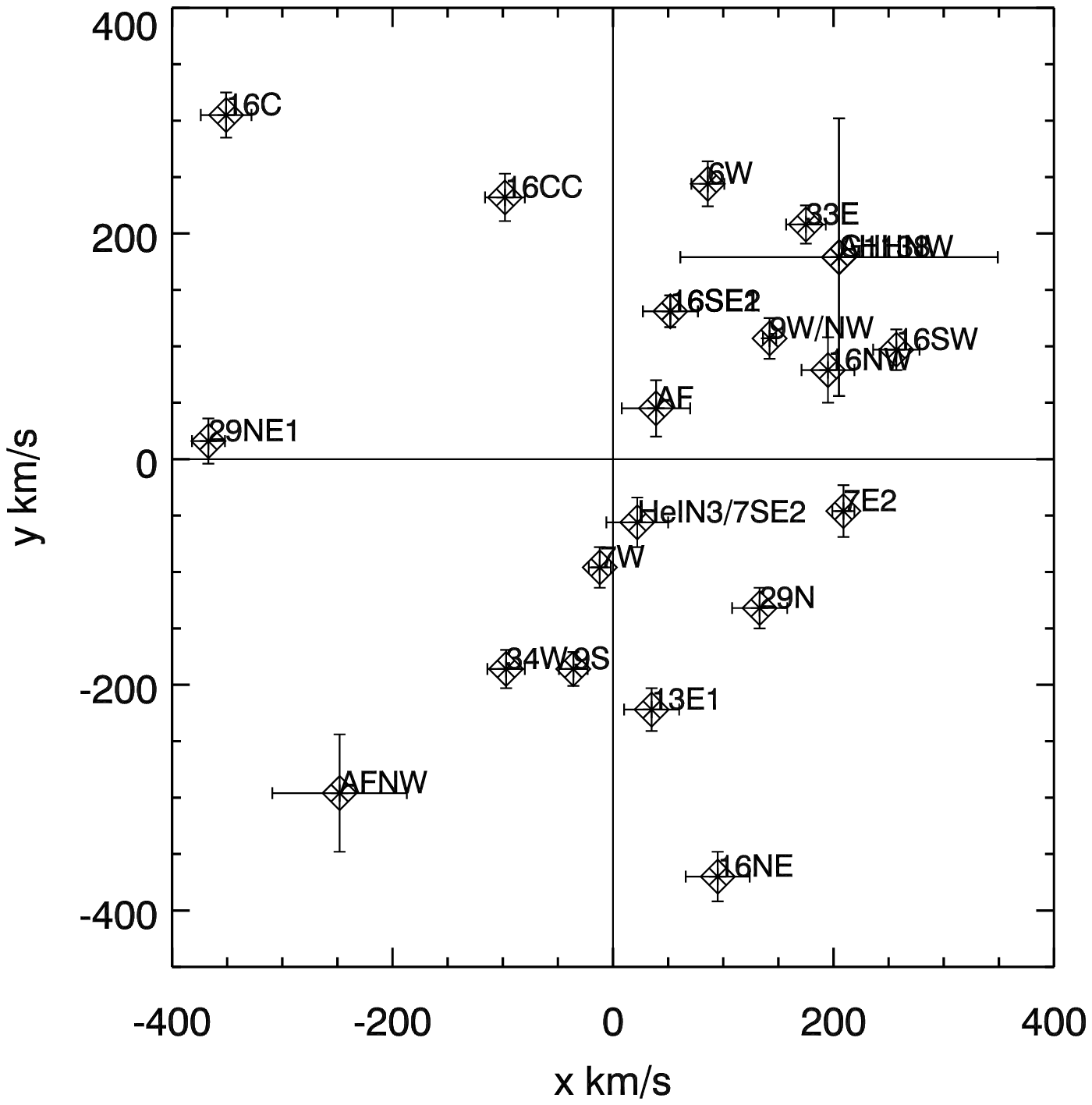}{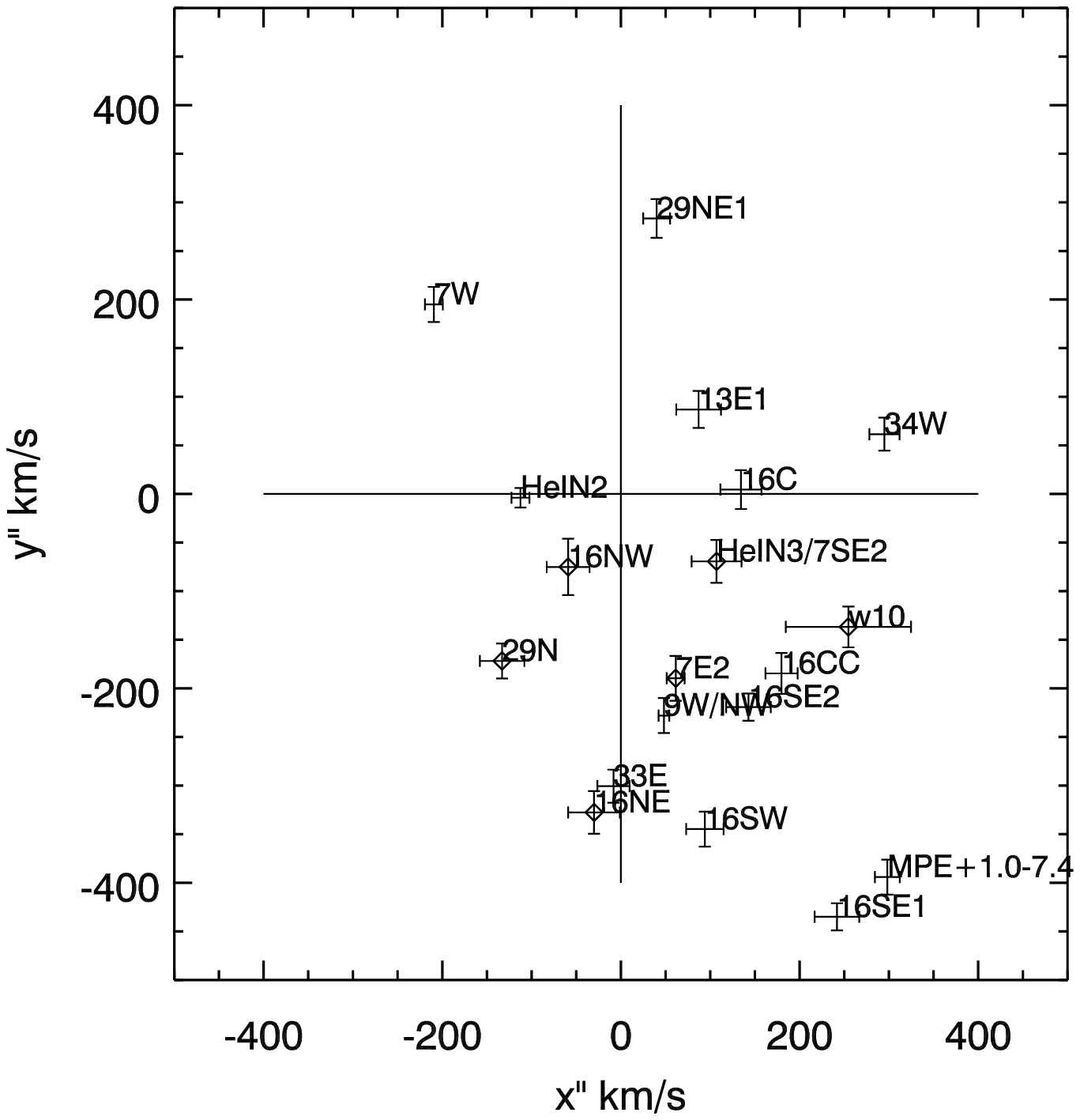}
\caption{{\it Left} - Plot of the velocities of the He I emission-line stars projected
on the plane of the sky, {\it Right} - Plot of the velocities of the stars in the frame
of reference within the plane of the derived disks. \label{diskplots}}
\end{figure}

The stars can be further divided into two separate populations of clockwise and counter-clockwise
orbital orientations through their normalized angular momentum along the line
of sight, J$_z$/J$_{z(max)}=(xv_y-yv_x)/(rv_r)$, the same method employed by Genzel et al. (2003). This term
is -1 when the orbit is counterclockwise tangential, 0 when its radial, and +1 when its clockwise
on the plane of the sky. 
When using our sample of 26 stars with published proper motion and radial velocity measurements, there
are 15 stars in the clockwise population and 11 stars in the counter-clockwise population corresponding
to positive and negative normalized angular momentum values, respectively (See Figure~\ref{jmaxplot}). 
Genzel et al. (2003) report 14 and 12 stars in these two populations, however, neither the
radial velocities nor the identities of these stars have been published to date. 
We estimate the orientations of the clockwise and counter-clockwise disks by minimizing the dot 
product of the unit normal vector and the stellar velocity vectors as done in 
Levin \& Beloborodov (2003) and Genzel et al. (2003). 
When the same stars are used with updated proper motion values from Ott (2004) and 
radial velocity values from this paper and 
the literature, we get a larger reduced chi squared value of 36 for the clockwise disk 
suggesting the need for a re-evaluation of the dynamics of these stars given the additional 
data and improved uncertainties. Also, IRS 29N is reported as a clockwise star in Levin \& Beloborodov (2003), however, this
may be an sign error since, when we use the proper motion and radial velocity for this star 
from Genzel et al. (2000),
we get the opposite result that it is a counter-clockwise orbiting star. 
When we include all the stars with published proper motion and radial velocities measurements 
shown in 
Figure~\ref{jmaxplot} with distances less than 7 arcseconds to reduce image distortion
effects to the proper motion values, the fits to the clockwise and counter-clockwise disks are 
poor with reduced 
chi-squared values of 30 for the 13 clockwise stars and 13 with 6 clockwise counter-clockwise stars. 
These chi-squared values can be compared to the 2 and 4.8 estimated for the 12 clockwise 
and 12 counter-clockwise stars, respectively, in Genzel et al. (2003) although we 
do not know which of our stars are in their sample.  
To improve these disk fits, we remove four
stars that appear to be outliers in velocity space when the
velocities are plotted from a frame of reference within the disk. 
This improves the fits to a small degree reduced chi squared values of 16 and 1.3
but with only ten and five stars remaining in the clockwise and
counter-clockwise disks, respectively. 
The normal vectors of these best fitting disks are 
${\bf n}$=[-0.50$\pm$0.06,-0.77$\pm$0.04,0.91$\pm$0.07] and 
${\bf n}$=[-0.18$\pm$0.1,-0.06$\pm$0.1,-0.77$\pm$0.1] and are consistent with 
the two disks derived in Genzel et al. (2003). Figure~\ref{diskplots} shows the
velocity of all the stars on the plane on the sky, as well as those stars used in the
best fitting disks as seen from an observer within the disk. While the stars do 
seem to fall along the vertical axis is this plot, there is still a
large scatter about the axis. When we derive a fit from the
combined sample of clockwise and counter-clockwise stars the reduced chi-squared value is 
42 suggesting that, while the two disk model is still a better fit to the data the
stars are not orbiting in as flat a disk as previously thought. 
 
\begin{figure}[ht]
\epsscale{1.0}
\plotone{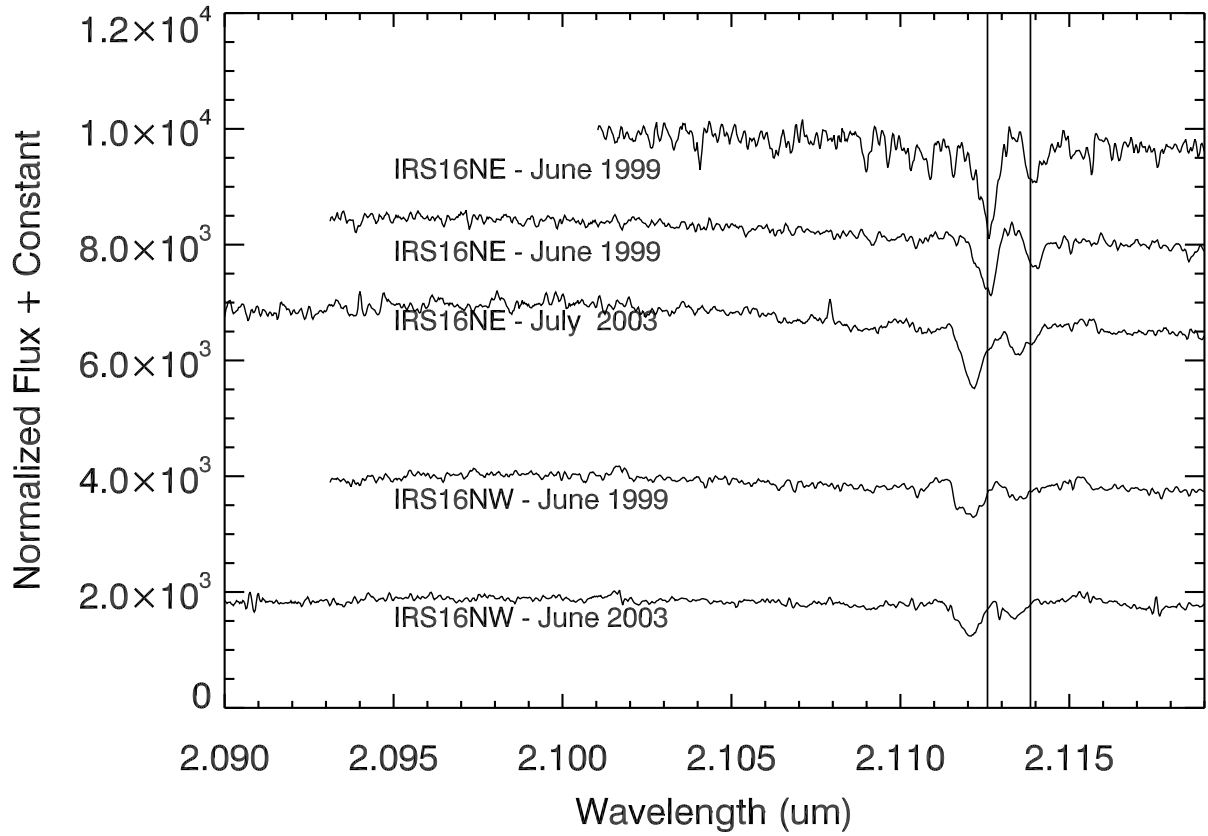}
\caption{Spectra of IRS 16NE and 16NW at different epochs. The vertical lines denote the 
estimated wavelength of the absorption features in the June/July 1999 epoch. A clear offset
can be seen in the position of the IRS 16NE lines in the June 2003 data with no such
offset apparent for IRS 16NW. \label{irs16spec}}
\end{figure}

Comparisons of the radial velocities estimated from NIRSPEC spectra of both
IRS 16NE and 16NW taken in 1999 and later in 2003 (see Figure~\ref{irs16spec}) reveal
an unexpected change of 60 \kms in the radial velocity of IRS 16NE over the five year time
span. IRS 16NE is three arcseconds away from Sgr A* compared to the 1.2 arcsecond separation
of IRS 16NW for which we detected no acceleration and use as a comparison to 16NE in
Figure~\ref{irs16spec}. However, the large error bars associated with the previous
measurements (See Table~\ref{irs16veltab}) of the radial velocity for this source do
not allow us to see any trends in the acceleration of the
source indicative of any orbital motion.
Ghez et al. (2004) report proper motions for IRS 16NE of 97$\pm$12 \kms East
and -383$\pm$19 \kms South with no detectable accelerations as large as that detected along the line of
sight. More likely, this star is a spectroscopic binary system which may have contributed to
the population of stars within one arcsecond of Sgr A* by ejecting stars
through gravitational interactions (Gould \& Quillen 2003).

\section{DISCUSSION}

This high-spatial resolution spectroscopic survey of the central
tens of arcseconds of the Galactic center has added five new
He I emission-line stars, an increase of almost 10\%, to the known population while
also reducing the uncertainties on the remaining previously published sources.
These five new sources include four narrow-line sources with He I $\lambda\lambda$ 2.112, 2.113 $\micron$
absorption features and one broad-line sources with wind broadened He I 2.058 $\micron$ emission
features. The spectra presented here are superior to those previously published due
to both higher spatial and spectral resolution. The high spectral resolution allows us to
detect both components of the He I $\lambda\lambda$ 2.112, 2.113 $\micron$ doublet which has not
been observed before in many of these sources. 

The occurrence of a few conflicting spectral 
classifications (IRS 29N and Blum WC9) shows the
importance of high spatial resolution observations to be able to distinguish between crowded sources
common in the Galactic center, as well as high spectral resolution observations to be able to discern
between photospheric and ISM features. Some of the earliest spectral classifications based on
low spatial resolution data were erroneous due to the misidentification
of Br $\gamma$ as being a photospheric line as opposed to originating from the ISM within
the mini-spiral (Krabbe et al. 1995). 
In order to create a complete understanding of the stellar population
within this region we must be able to separate the mini-spiral emission from that
of the stars. Reproducing the spectral features of the mini-spiral based on its known
kinematics and density could then be used to separate emission from the gas and the stars. 

Paumard et al. (2001, 2003) use the narrow FWHM of the 2.058 $\micron$
line as well as the large K magnitudes of the IRS 16 sources to claim they are LBV-like stars. 
This, however, did not consider whether the stars show any sign
of variability, nor do they have spectra resembling those associated with known
LBV's (Humphreys \& Davidson 1994; Figer et al. 1998). Ott et al. (1999) 
provided light curves of IRS 16C, 16CC, 16NW, and 33E showing
no sign of variability. Finally, it would be surprising, indeed if these stars were
all {\it bona fide} LBVs given their short lifetimes and relative paucity in the rest of the galaxy.
We conclude that the narrow-line stars, with the exception of IRS 16NE, more
closely resemble B supergiants based on the equivalent width of their He I 
$\lambda\lambda$ 2.112, 2.113 $\micron$ lines, the absence of any detectible NIII 2.1155 $\micron$
emission and their similarity to such stars identified in the Quintuplet (Hansen et al. 1996; Figer et
al. 1999).
Despite our determination that the narrow-line sources have spectral
types near early B stars, we cannot trace the precursor ($>$20 M$_{\sun}$) stars back
to a precise epoch of star formation since stars in this phase race back and forth on
the Hurtzsprung-Russell diagram. The same ambiguities exist for estimating an exact mass for
these stars since the precursor mass for a WC star can range from 25 to 120 M$_{\sun}$ (Abbott \& Conti 1987).

Table~\ref{poptab} provides a list of the number of OB supergiants (OBI), early transitional
objects including LBV and sgB[e], red supergiants (RSG), and WN and WC Wolf-Rayet stars 
identified in the Central cluster, the nearby Arches and Quintuplet clusters and the young galactic
open cluster, Westerlund 1 (Wd 1) (Clark et al. 2005; Figer et al. 2004 and reference within). As
we are improving our knowledge of the stellar populations of these clusters, it appears that
Central and Quintuplet clusters have similar distributions of spectral types which is not
surprising given their similar cluster mass and age (Figer et al. 2004). The larger number of
Wolf-Rayet stars in the Central cluster compared to Wd 1 might be due to the different environments
in which these two clusters of similar age formed (Clark et al. 2004). 

As mentioned in Section 4, our radial velocity results marginally agree with the
hypothesis that the He I emission-line stars are orbiting coherently in either a 
clockwise and counter-clockwise orbiting disk. 
These results are based on measurements of just four out of six of 
the phase-space coordinates (x, y, v$_x$ and v$_y$).
While we have increased the number of stars with radial velocity
measurements, their line-of-sight distance from Sgr A* is unknown. 
Those sources embedded within the Northern Arm overcome this issue by utilizing the
three-dimensional models of the Northern Arm by Paumard et al. (2003b)
to deduce their line-of-sight distance. The orbital parameters of the Northern
Arm sources, IRS 1W and 10W, estimated from all six phase space
coordinates, are consistent with those of the clockwise orbiting disk, supporting
the existence of at least one of the two disks. The improved uncertainties
for both the proper-motion and radial velocity measurements result in $\chi^2$ fits with
larger reduced chi-squared values than those previously published suggesting that the
disks may not be as coherent as previously thought. If the orbits of the He I emission-line stars
do comprise one or more coherent structures, then this lends credence to the
hypothesis that the young star population is the result of the in-fall of a massive 
star cluster. In fact, it has been recently suggested that the nearby compact
cluster of stars known as IRS 13 could be the remnants of the core of such
a cluster (Maillard et al. 2004). However, the cluster in-fall models of Kim and Morris (2003)  
predict up to a factor of ten too many He I emission-line stars in the central few arcseconds so
our increase in the known population of these stars does not help this discrepancy.

Because of their high mass-loss rates, the population of He I emission-line stars has been hypothesized to
be a significant source of accretion of material onto Sgr A*  (Quataert et al. 2003; Melia et al. 1999). 
However, the bulk of the accretion comes from the copious mass-loss of the
IRS 13 sources. When we estimate the additional contributions from our newly
identified sources to the effective accretion rate onto Sgr A*, despite increasing
the population, there is not a significant additional source of material
accreting onto Sgr A*.

\section{CONCLUSIONS}

We have added five new \ion{He}{1} emission-line stars to the known population of hot, 
massive stars in the central parsec of the galaxy and have confirmed 
the stellar classification of an additional fourteen stars. These stars are
joined by two others in our sample which have radial velocities not published to
date. With improved uncertainties for both the proper-motions and radial velocities of these
stars, we somewhat agree with the hypothesis that the stars form two coherent disks
although the degree of coherence is still under speculation as our fits to the two disk model are
not all that satisfactory and are in a different orientation than that previously 
published. The addition of more He I emission-line stars to the sample and improved 
uncertainties on the proper motions of the stars further from Sgr A* will help
come to a firm conclusion on the dynamics of these stars. These results are important
for testing the origin of these young, massive stars near the supermassive black hole
since this large population is still
difficult to explain, given the dynamic environment of the Galactic center.  
Finally, we have found preliminary evidence that the He I
emission-line star, IRS 16NE, may be a spectroscopic binary based on an observed change in its
radial velocity over five years.

More conclusive classifications of the spectral types of these stars through
careful study of their spectral features will allow us to derive a
comprehensive census of these stars to also compare to the star formation
models which predict the expected number stars of specific spectral types.
Integral field spectrographs will greatly benefit this field as they enable deeper
observations to begin to probe the main sequence. It will also be of interest to extend the wavelength 
coverage to longer wavelengths such as the L or L' bands so that the additional
spectral information provides more robust classifications.

\begin{acknowledgements}

Data presented herein were obtained at the W.M. Keck Observatory, 
which is operated as a scientific partnership among the California Institute of Technology, 
the University of California and the National Aeronautics and Space Administration. 
The Observatory was made possible by the generous financial support of the W.M. Keck Foundation.
The authors wish to recognize and acknowledge the very significant cultural role and 
reverence that the summit of Mauna Kea has always had within the 
indigenous Hawaiian community.  
We are most fortunate to have the opportunity to conduct observations from this mountain.

We also acknowledge the work of: Maryanne Angliongto, Oddvar
Bendiksen, George Brims, Leah Buchholz, John Canfield, Kim Chin, Jonah
Hare, Fred Lacayanga, Samuel B. Larson, Tim Liu, Nick Magnone, Gunnar
Skulason, Michael Spencer, Jason Weiss and Woon Wong. In addition, we
thank the CARA instrument specialist Thomas A. Bida, and all the CARA staff
involved in the commissioning and integration of NIRSPEC. F.N. acknowledges PNAYA2003-02785-E and AYA2004-08271-C02-02 grants and
the Ramon y Cajal program.

\end{acknowledgements}

\begin{deluxetable}{lcc|lcc}
\tablecaption{Line List \label{linetab}}
\footnotesize
\tablewidth{30pc}
\tablehead{
\colhead{Transition} & \colhead{$\lambda$ [$\micron$]} & \colhead{Origin} & \colhead{Transition} & \colhead{$\lambda$ [$\micron$]} & \colhead{Origin}
}
\startdata
\ion{He}{1}   & 2.0378  & stellar    & OH          & 2.1068 & Telluric \\
OH            & 2.0413 & Telluric    & OH          & 2.1097 & Telluric\\
OH            & 2.0499 & Telluric    & \ion{C}{3}  & 2.1038 & Stellar\\
\ion{He}{1}   & 2.0422 & Stellar     & \ion{C}{3}  & 2.1081 & Stellar\\
OH            & 2.0564 & Telluric    & OH          & 2.1106 & Telluric \\
\ion{He}{1}   & 2.0582 & Stellar     & \ion{He}{1} & 2.1126 & Stellar\\
\ion{He}{1}   & 2.0587 & Stellar/ISM & \ion{He}{1} & 2.1137 & Stellar\\
OH            & 2.0673 & Telluric   & \ion{C}{3}   & 2.1152 & Stellar\\
OH            & 2.0729 & Telluric   & \ion{C}{3}   & 2.1155 & Stellar\\
\ion{C}{4}    & 2.0705 & Stellar     & \ion{C}{3}  & 2.1156 & Stellar\\
\ion{He}{1}   & 2.0762 & Stellar     & OH          & 2.1116 & Telluric \\
\ion{C}{4}    & 2.0796 & Stellar     & OH          & 2.1156 & Telluric  \\
\ion{C}{4}    & 2.0842 & Stellar     & OH          & 2.1177 & Telluric \\
OH            & 2.0860 & Telluric   & \ion{He}{1}  & 2.1203 & Stellar\\
OH            & 2.0910 & Telluric   & \ion{C}{3}   & 2.1217 & Stellar\\
\ion{He}{1}   & 2.0968 & Stellar     & OH          & 2.1233 & Telluric \\
\ion{N}{5}    & 2.0997 & stellar     & H$_2$       & 2.1218 & ISM \\
\enddata
\end{deluxetable}

\begin{deluxetable}{lccc}
\tablecaption{Identified Lines in Source Spectra \label{lineident}}
\tablewidth{30pc}
\tablehead{
\colhead{Source} & \colhead{\ion{He}{1} 2.058} & \colhead{\ion{He}{1} 2.112/2.113} & \colhead{\ion{He}{1} 2.166$^a$}\\
\colhead{}       & \colhead{$\micron$}   & \colhead{$\micron$}        & \colhead{$\micron$}
}
\startdata
IRS 16NE & P-Cygni?  & abs & P-Cygni \\
IRS 16NW & em       & abs & P-Cygni\\
IRS 16C  & P-Cygni?  & abs & P-Cygni\\
IRS 16CC & em       & abs & P-Cygni?\\
IRS 16SW(W) & em+abs   & abs &  em\\
G1438    & ... & abs & ... \\
IRS 33E  & em & abs & strong em\\
IRS 33N/A11  & em & abs & strong em \\
IRS 6W   & em & abs & strong em\\
Blum WC9 & em & abs & weak em\\
A13  & ... & abs & weak em\\
A12      & ... & abs & em\\
-9.13-2.14(248)$^b$ & em & abs & em \\
\hline
AF       & broad P-Cygni& broad P-Cygni& broad em\\
AFNW     & broad P-Cygni& broad em& broad em\\
AFNWB    & broad em& broad em&  em\\
IRS 15SW & broad em& broad em& broad em\\
IRS 7W   & broad P-Cygni& broad em & broad em\\
IRS 9S   & em & broad em& em\\
G1138    & em & broad em&  em \\
GCHe2       & em & broad em&  em\\
\enddata
\tablenotetext{a}{Any emission lines observed at this wavelength most probably originate from the diffuse ISM within the mini-spiral.}
\tablenotetext{b}{This source did not have names in the literature, therefore, we
assigned them these designations based on their offset in arcseconds with respect to Sgr A*. The number in
parentheses corresponds to its place in our data list and can be used to find it in Figure~\ref{image}.}
\end{deluxetable}

\begin{deluxetable}{lcccc}
\tablewidth{28pc}
\tablecaption{Offsets$^a$ and Proper motions of the \ion{He}{1} $\lambda\lambda$ 2.112,2.113 $\micron$ Absorption Doublet Stars \label{nartab1}}
\footnotesize
\tablehead{
\colhead{}       & \colhead{$\Delta RA$} &\colhead{$\Delta Dec$} &
\colhead{PM$_\alpha$}  & \colhead{PM$_\delta$}   \\
\colhead{Source} & \colhead{Arcseconds}     & \colhead{Arcseconds} & \colhead{\kms} & \colhead{\kms}
}
\startdata
IRS 16NE   &2.835$\pm$0.007&1.058$\pm$0.005 & 95$\pm$29 &-370$\pm$22   \\
IRS 16NW   &-0.028$\pm$0.006&1.242$\pm$0.007 & 195$\pm$24 &  79$\pm$29   \\
IRS 16C    &1.182$\pm$0.006&0.509$\pm$0.027 &-351$\pm$23  & 305$\pm$20   \\
IRS 16CC   &1.963$\pm$0.004&0.563$\pm$0.005 & -98$\pm$18 & 232$\pm$21  \\
IRS 16SW(W) & 1.007$\pm$0.005&-0.952$\pm$0.004& 257$\pm$21&97$\pm$18  \\
IRS 33E    &0.620$\pm$0.005&-3.107$\pm$0.004 & 175$\pm$18 & 208$\pm$17\\
IRS 6W     &-5.273$\pm$0.004   &0.804$\pm$0.005 & 86$\pm$15  & 244$\pm$20  \\
IRS 33N/A11    &-0.096$\pm$0.004&-2.166$\pm$0.005 & 48$\pm$17 & -172$\pm$18   \\
BlumWC9    &  -8.418           & -6.382           &            &           \\
-9.13-2.14 &  -9.133      &   -2.135        &            &            \\
G1438$^b$      &   -7.805        &    -8.494      &            &             \\
A12        & -0.456 &  8.487         &            &               \\
A13        & -0.612 &   10.098              &            &   \\
\enddata
\tablenotetext{a}{Offsets are with respect to Sgr A*. Those sources with both offsets
and proper motions are from Ott (2004), otherwise the offsets are estimated from our
NIRSPEC images.}
\tablenotetext{b}{The nomenclature for this star comes from Figer et al. (2003)}
\end{deluxetable}

\begin{deluxetable}{lccccccccc}
\rotate
\tablecaption{Radial Velocities and EWs Derived from the \ion{He}{1} $\lambda\lambda$ 2.112,2.113 $\micron$ Absorption Doublet \label{nartab2}}
\footnotesize
\tablehead{
\colhead{}       & \multicolumn{2}{c}{June} &  \multicolumn{2}{c}{July}   & \colhead{}  & \colhead{} & & \colhead{P2001$^a$}        &  \colhead{G2000}   \\
\colhead{Source} & \colhead{Vr$_{2.112}$}  & \colhead{Vr$_{2.113}$} &  \colhead{Vr$_{2.112}$}& \colhead{Vr$_{2.113}$}  & \colhead{Vr$_{avg}$   }& \colhead{$\sigma$}     &  \colhead{EW}   & \colhead{Vr$_{2.058}$} &  \colhead{Vr$_{2.058}$} \\
                 & \colhead{\kms}   &  \colhead{\kms} &   \colhead{\kms} &  \colhead{\kms} &  \colhead{\kms} &  \colhead{\kms} &  \colhead{$\AA$} & \colhead{\kms} & \colhead{\kms}
}
\startdata
IRS 16NE   &   4 &  36 & -8  &  19   & 12  & 19 &0.97  &1$^{+35}_{-21}$    & 17$\pm$25 \\
IRS 16NW   & -65 & -28 & -64 & -40 & -49 & 18   &1.41  &-46$^{+29}_{-31}$  & -30$\pm$30 \\
IRS 16C    & 108 & 127 &  98 & 131  & 116 & 15  &1.26  &-24$^{+17}_{-16}$  & -180$\pm$25 \\
IRS 16CC   & 252 & 263 & 226 & 243 & 246 & 15   &1.84  &P2003              & 245$\pm$70\\
IRS16SW(W)$^b$ & 242 & 252 &     &   & 247 & 5  &1.66  &P2003              & 460$\pm$30 \\
IRS 33E    & 121 & 151 &     &        & 136 &15 &1.93  &258$_{-84}^{+57}$  & 160$\pm$60\\
IRS 6W     & 63  & 62  &     &      & 63 & 1    &1.88  &                   & -150$\pm$70 \\
BlumWC9$^c$  & 88  & 79  &  &  & 83& 4          &1.34  &P2003              &  \\
-9.13-2.14$^{\dagger\dagger}$&  42  & 37  & -8  & 19  & 22 & 22&1.35&                      &  \\
G1438$^{\dagger\dagger}$      &  139 & 157 &     &  & 148 &9       &0.56&                      &  \\
IRS 33N/A11 & 50  &  61 &     & & 55 & 5    &2.40&       P2004        &  \\
A12$^{\dagger\dagger}$ & -148 &-150 &  &    & -149 & 1    &1.29&                      &  \\
A13$^{\dagger\dagger}$& -233 &-220 & -238&-250&-235&12 &1.16&                      &  \\
\enddata
\tablenotetext{a}{Those Paumard radial velocities noted by "P2003" are listed in this paper as being
\ion{He}{1} emission-line stars but with no values for the radial velocities provided.}
\tablenotetext{b}{IRS 16SW is a known eclipsing binary (Ott et al. 1999) resulting in a
bias in the value of its radial velocity due to its non-negligible line-of-sight
orbital velocity.}
\tablenotetext{c}{Blum WC9 is classified as a broad-line source by Paumard et al. (2003).}
\tablenotetext{\dagger\dagger}{New He I emission line stars and radial velocity detections}
\tablenotetext{\dagger}{New radial velocity detections}
\end{deluxetable}

\begin{deluxetable}{lcccc}
\tablewidth{25pc}
\tablecaption{Offsets and Proper Motions for the \ion{He}{1} 2.058$\micron$ Broad Emission Line Stars\label{brotab1}}
\footnotesize
\tablehead{
\colhead{}        &\colhead{$\Delta$ RA$^a$}& \colhead{$\Delta$ Dec} &\colhead{V$_\alpha$}&\colhead{V$_\delta$}\\
\colhead{Source}  &\colhead{Arcseconds} & \colhead{Arcseconds}   &\colhead{\kms} &\colhead{\kms}
}
\startdata
AF         &-6.541$\pm$0.008 & -6.919$\pm$0.006  &39$\pm$31    &45$\pm$25   \\
AFNW       &-7.699$\pm$0.015 & -3.541$\pm$0.013  &-248$\pm$61  &-296$\pm$52  \\
AFNWB      &-8.177$\pm$0.004 &-2.820$\pm$0.005   &205$\pm$144  &179$\pm$123     \\
IRS 7W     &-4.049$\pm$0.002 & 4.980$\pm$0.004   & -12$\pm$10  & -96$\pm$18\\
IRS 15SW   & -1.50            &10.10              &             &            \\
GCHe2      & 2.847$\pm$0.005 &-5.620$\pm$0.005   &142$\pm$6    &107$\pm$18 \\
G1138$^b$      &   -7.873        &-6.712             &             &             \\
IRS 9S     &5.683$\pm$0.004 &-8.200$\pm$0.005 &-36$\pm$13  &-186$\pm$15    \\
\enddata
\tablenotetext{a}{Offsets are with respect to Sgr A*. Those sources with both offsets
and proper motions are from Ott (2004), otherwise the offsets are estimated from our
NIRSPEC images.}
\tablenotetext{b}{The nomenclature for this star comes from Figer et al. (2003)}
\end{deluxetable}

\begin{deluxetable}{lcccccccc}
\rotate
\tablecaption{Radial Velocities Derived from the \ion{He}{1} 2.058$\micron$ Broad Emission Line Stars\label{brotab2}}
\footnotesize
\tablehead{
\colhead{} &\multicolumn{2}{c}{June} &  & \colhead{V$_{avg}$} & \colhead{$\sigma$} & \multicolumn{2}{c}{P2001} & \colhead{G2000} \\
\colhead{} &\colhead{Vr$_{2.058}$ Fit} &\colhead{FWHM} & \colhead{Vr$_{2.058}$ CC} & & \colhead{Vr$_{2.058}$}  & \colhead{Vr$_{2.058}$} & \colhead{FWHM} & \colhead{Vr$_{2.058}$} \\
\colhead{Source}  &\colhead{\kms} & \colhead{\kms}  &  \colhead{\kms} & \colhead{\kms} &\colhead{\kms} & \colhead{\kms} & \colhead{\kms}  & \colhead{\kms}
}
\startdata
AF            & 248   & 749     & 265 & 257  & 10 &  192$^{+52}_{-64}$   &  826  & 140$\pm$50  \\
AFNW          & 206   & 1324    & 187 & 197  & 10 &  159$^{+84}_{-79}$   & 1300  & 150$\pm$70 \\
AFNWB$^\dagger$     & 218   & 1690    & 283 & 251  & 30 &                      &       &            \\
IRS 7W        & -346  & 811     & -207 & -277 & 70 &   -292$^{+49}_{-41}$ & 1031  & -300$\pm$50\\
IRS 15SW      & -204  & 790     & -124 & -164& 40 &   -179$^{+24}_{-26}$ & 894   & -230$\pm$50 \\
GCHe2$^\dagger$     & 108   & 844     & 194 & 151 & 40 &                      &       &          \\
G1138$^{\dagger\dagger}$     &   212 & 1677    & 150 & 181 & 30 &                      &       &          \\
IRS 9S        &   180 & 1321    & 146 & 163 & 20 &                      &       &  200$\pm$50 \\
\enddata
\tablenotetext{\dagger\dagger}{New He I emission line stars and radial velocity detections}
\tablenotetext{\dagger}{New radial velocity detections.}
\end{deluxetable}

\begin{deluxetable}{lcccccccc}
\rotate
\tablecaption{Properties of \ion{He}{1} emission lines discovered in other surveys\label{remtab}}
\footnotesize
\tablehead{
                  &\colhead{$\Delta$ RA$^a$}& \colhead{$\Delta$ Dec} &\colhead{V$_\alpha$} & \colhead{V$_\delta$}& \colhead{V$_r$(P2001)$^b$} & \colhead{V$_r$(G2000)} & Type \\
\colhead{Source}  &\colhead{Arcseconds} &\colhead{Arcseconds}    &\colhead{\kms}& \colhead{\kms} & \colhead{\kms} & \colhead{\kms}
}
\startdata
IRS 34W       &-4.132$\pm$0.004 & 1.646$\pm$0.004 & -97$\pm$17   & -186$\pm$17 & -175$^{+117}_{-94}$   & -215$\pm$30 &Narrow\\
ID180         &9.474$\pm$0.001  &0.58$\pm$0.017   &-73$\pm$210   &-10$\pm$80   & 198$^{+130}_{-114}$   &  &  Broad\\
He1N3         & 3.01            &3.52             &              &             & 615$^{+133}_{-152}$   & & Broad\\
IRS 7E2       &4.394$\pm$0.005  &4.976$\pm$0.004  &  209$\pm$10  &-46$\pm$23   & P2003                 & & Broad \\
IRS 29N       &-1.666$\pm$0.004 & 1.456$\pm$0.004 & 181$\pm$10   & -129$\pm$10 & P2003                 & & Broad \\
IRS 15NE      &1.60             &11.40            &              &             & P2003                 & -80$\pm$50      & Broad\\
IRS 16SE2/35W &2.904$\pm$0.004  &-1.170$\pm$0.004 &52$\pm$25     &131$\pm$14   & P2003                 & 265$\pm$90 & Broad\\
W10           &-1.419$\pm$0.004 &-0.270$\pm$0.005 & -124$\pm$21  & -235$\pm$19 &                       &  270$\pm$70 &  \\
IRS 16SE1     & 1.869$\pm$0.001 &-1.071$\pm$0.001 & 174$\pm$10   & 102$\pm$10  &                       & 450$\pm$60  & \\
IRS 29NE1     &-0.91$\pm$0.004  & 2.04$\pm$0.005  &-367$\pm$15   & 16$\pm$20   &                       & -130$\pm$100 & \\
IRS 7SE       & 2.55            & 3.19            & 309$\pm$159  & 429$\pm$105 &                       & -85$\pm$40  & \\
IRS 7E        & 4.314$\pm$0.004 & 5.328$\pm$0.005 & -18$\pm$12   &-21$\pm$19   &                       & -20$\pm$50 &  \\
AFNWW         &-9.15$\pm$0.005  &-3.30$\pm$0.005  &              &             &                       & 250$\pm$100  & \\
IRS 1S        &5.55             &0.45             &              &             &                       & -300$\pm$200 & \\
IRS 13E$^c$   &-3.24            &-1.68            &              &             &                       & &Broad \\
\enddata
\tablenotetext{a}{Proper motions and offsets from Ott, PhD 2004, otherwise they are from P2001 or G2000.}
\tablenotetext{b}{Those Paumard radial velocities noted by "P2003" are listed in Paumard et al. (2003) as being
\ion{He}{1} emission-line stars but with no values for the radial velocities provided.}
\tablenotetext{c}{The IRS 13E cluster has now been resolved into seven 
separate components whose offsets and proper motions
are provided in Maillard et al. (2004). The offset given here is for IRS 13E2.}
\end{deluxetable}

\begin{deluxetable}{llll}
\tablecaption{Hot stars with no photospheric lines \label{nondet}}
\tablewidth{40pc}
\footnotesize
\tablehead{
\colhead{Star} & \colhead{Comment} & \colhead{Star} & \colhead{Comment}
}
\startdata
IRS 5       & NA source                    & G803  & \\
IRS 26      &                              &  G604A        &\\
IRS 10W     & NA source                    & IRS 29NE1    &  \\
IRS 3       & featureless                  & W10          & G2000 \\
A22/OSUHe1  &                              & IRS 35       & \\
IRS 29NE1   & G2000                        & G345         & \\
IRS 29N     & WC9/Broad star (P2001)       & AFNWW & G2000 \\
IRS 1W      &  NA source                   & IRS 9NW      & \\
G796        &                              & 16SE2        & WN5/6 star (Horrobin et al. 2004)\\
IRS 34W     & Narrow line star (P2001)     & +10.82+7.27    & \\
MPE+1.6-6.8 &                              & -6.42+4.06     & \\
IRS 35W     & Broad line star (P2003)      & +11.04+3.92    & \\
IRS 6E      & WC9 star Krabbe et al. 1995  & +9.70+3.58     &\\
IRS 21      & NA source                    & -8.05+2.22     & \\
MPE-2.0-8.5 & center of mini-cavity        & -3.00+1.27     &\\
IRS 2       & NA source                    & -5.36-1.82     & \\
A21         & center of Eastern cavity     & -5.14-1.94      &  \\
IRS 20      &                              & -7.08-2.26     & \\
MPE+1.4-12.2 &                             &  -10.03-4.49   & \\
IRS 14S      &                             & +5.40-5.52     & \\
A22          &                             &                & \\
124          &                             &                & \\
\enddata
\end{deluxetable}

\begin{deluxetable}{lcc}
\tablecaption{Measured Radial Velocities of IRS 16NE \label{irs16veltab}}
\tablewidth{23pc}
\tablehead{
\colhead{Epoch} & \colhead{V$_z$} & \colhead{Reference} \\
\colhead{}      & \colhead{\kms} & \colhead{}
}
\startdata
Spring 1996 & 17$\pm$25 & Genzel et al. (2000) \\
July 25 1997 & 1$\pm$25 & Paumard et al. (2001) \\
June 4 1999 & 4$\pm$16 & this paper \\
June 21 2003 & -52$\pm$11 & this paper \\
\enddata 
\end{deluxetable}

\begin{deluxetable}{lccccc}
\tablecaption{Stellar Populations for Known Massive Clusters \label{poptab}}
\tablewidth{27pc}
\tablehead{
\colhead{Cluster} & \colhead{OB1} & \colhead{Early Trans} & \colhead{RSG} & \colhead{WN} & \colhead{WC}\\
}
\startdata
Quintuplet$^{a}$  & 14    & 2    & 1 & 5     & 11 \\
Arches$^{a}$      & 20    & 0    & 0 & $>$6  & 0  \\
Center$^{a}$      & $>$10 & 1    & 2 & $>$10 & $>$15  \\
Wd 1$^{b}$        & $>$25 & $>$5 & 3 & $>$7  & $>$6 \\
\enddata 
\tablenotetext{a}{Taken from this work, Figer et al. (2004) and reference within}
\tablenotetext{b}{Taken from Clark et a. 2005}
\end{deluxetable}

\clearpage 

\newpage
\clearpage
\clearpage
\clearpage
\clearpage
\clearpage
\clearpage
\clearpage
\clearpage
\clearpage
\clearpage
\clearpage

\newpage

\end{document}